\documentclass[manuscript]{aastex631}

\usepackage{amsmath}
\usepackage{verbatim}
\usepackage{hyperref}

\newcommand{\pivec}{\mbox{\boldmath $\pi_{\rm E}$}}
\newcommand{\pieE}{\mbox{$\pi_{{\rm E},E}$}}
\newcommand{\pieN}{\mbox{$\pi_{{\rm E},N}$}}

\newcommand{\fifteenninetyeight}{OGLE-2016-BLG-1598}    % KMT-2016-BLG-0696    OGLE-2016-BLG-1598(*)  MOA-2016-BLG-521
\newcommand{\eighteendoubleO}{OGLE-2016-BLG-1800}       % KMT-2016-BLG-0781    OGLE-2016-BLG-1800(*)  MOA-2016-BLG-581
\newcommand{\fivetwentysix}{MOA-2016-BLG-526}           % KMT-2016-BLG-1611    OGLE-2016-BLG-1705     MOA-2016-BLG-526(*)
\newcommand{\twentythreetwentyone}{KMT-2016-BLG-2321}   % KMT-2016-BLG-2321    --                     --
\newcommand{\twelvefourtythree}{KMT-2016-BLG-1243}      % KMT-2016-BLG-1243    --                     --
\newcommand{\threethirtysix}{OGLE-2016-BLG-0336}        % KMT-2016-BLG-1406    OGLE-2016-BLG-0336(*)  MOA-2016-BLG-092
\newcommand{\eighteightytwo}{OGLE-2016-BLG-0882}        % KMT-2016-BLG-1449    OGLE-2016-BLG-0882(*)  --
\newcommand{\seventeenfour}{OGLE-2016-BLG-1704}         % KMT-2016-BLG-1609    OGLE-2016-BLG-1704(*)  --
\newcommand{\fourteeneight}{OGLE-2016-BLG-1408}         % KMT-2016-BLG-1630    OGLE-2016-BLG-1408(*)  --
\newcommand{\twentythreeninetynine}{KMT-2016-BLG-2399}  % KMT-2016-BLG-2399    --                     --
\newcommand{\twentyfourseventythree}{KMT-2016-BLG-2473} % KMT-2016-BLG-2473    --                     --
\newcommand{\sixtwenty}{OGLE-2016-BLG-0620}             % KMT-2016-BLG-0255    OGLE-2016-BLG-0620(*)  MOA-2016-BLG-183
\newcommand{\ninethirteen}{KMT-2016-BLG-0913}           % KMT-2016-BLG-0913    --                     --
\newcommand{\fourteenthirtytwo}{OGLE-2016-BLG-1432}     % KMT-2016-BLG-1004    OGLE-2016-BLG-1432(*)  --
\newcommand{\twelvetwentytwo}{KMT-2016-BLG-1222}        % KMT-2016-BLG-1222    --                     --
\newcommand{\eighteenfourtyfour}{OGLE-2016-BLG-1844}    % KMT-2016-BLG-1326    OGLE-2016-BLG-1844(*)  --
\newcommand{\fourteentwentyfive}{KMT-2016-BLG-1425}     % KMT-2016-BLG-1425    --                     --
\newcommand{\nineeightytwo}{OGLE-2016-BLG-0982}         % KMT-2016-BLG-1433    OGLE-2016-BLG-0982(*)  --
\newcommand{\fifteenseventeen}{OGLE-2016-BLG-1517}      % KMT-2016-BLG-1461    OGLE-2016-BLG-1517(*)  --
\newcommand{\twelvefiftyeight}{OGLE-2016-BLG-1258}      % KMT-2016-BLG-2067    OGLE-2016-BLG-1258(*)  --
\newcommand{\twentytwofiftysix}{KMT-2016-BLG-2256}      % KMT-2016-BLG-2256    --                     --
\newcommand{\twentythreethirtyone}{KMT-2016-BLG-2331}   % KMT-2016-BLG-2331    --                     --
\begin{document}

\title{Systematic KMTNet Planetary Anomaly Search. XI. Complete Sample of 2016 Sub-Prime Field Planets
}

% Author List ------------------------------------------------------------------------------------------------------------
% 1
\author{In-Gu Shin} 
\affiliation{Center for Astrophysics $|$ Harvard \& Smithsonian 60 Garden St., Cambridge, MA 02138, USA}
%\correspondingauthor{In-Gu~Shin} \email{ingushin@gmail.com}
% 2
\author{Jennifer C. Yee}
\affiliation{Center for Astrophysics $|$ Harvard \& Smithsonian 60 Garden St., Cambridge, MA 02138, USA}
% 3
\author{Weicheng Zang}
\affiliation{Center for Astrophysics $|$ Harvard \& Smithsonian 60 Garden St., Cambridge, MA 02138, USA}
\affiliation{Department of Astronomy, Tsinghua University, Beijing 100084, China}
% 4
\author{Cheongho Han}
\affiliation{Department of Physics, Chungbuk National University, Cheongju 28644, Republic of Korea}
% 5
\author{Hongjing Yang}
\affiliation{Department of Astronomy, Tsinghua University, Beijing 100084, China}
% 6
\author{Andrew Gould}
\affiliation{Max Planck Institute for Astronomy, K\"onigstuhl 17, D-69117 Heidelberg, Germany}
\affiliation{Department of Astronomy, The Ohio State University, 140 W. 18th Ave., Columbus, OH 43210, USA}
% 7
\author{Chung-Uk Lee}
\affiliation{Korea Astronomy and Space Science Institute, Daejeon 34055, Republic of Korea}
% 8
\author{Andrzej Udalski}
\affiliation{Astronomical Observatory, University of Warsaw, Al.~Ujazdowskie 4, 00-478 Warszawa, Poland}
% 9
\author{Takahiro Sumi}
\affiliation{Department of Earth and Space Science, Graduate School of Science, Osaka University, Toyonaka, Osaka 560-0043, Japan}
\collaboration{10}{(Leading authors),}
%
% KMTNet ---------------------------
% Science Team
% 1
\author{Michael D. Albrow} 
\affiliation{University of Canterbury, Department of Physics and Astronomy, Private Bag 4800, Christchurch 8020, New Zealand}
% 2
\author{Sun-Ju Chung}
\affiliation{Korea Astronomy and Space Science Institute, Daejeon 34055, Republic of Korea}
% 3
\author{Kyu-Ha Hwang}
\affiliation{Korea Astronomy and Space Science Institute, Daejeon 34055, Republic of Korea}
% 4
\author{Youn Kil Jung}
\affiliation{Korea Astronomy and Space Science Institute, Daejeon 34055, Republic of Korea}
\affiliation{Korea University of Science and Technology (UST), 217 Gajeong-ro, Yuseong-gu, Daejeon 34113, Republic of Korea}
% 5
\author{Yoon-Hyun Ryu}
\affiliation{Korea Astronomy and Space Science Institute, Daejeon 34055, Republic of Korea}
% 6
\author{Yossi Shvartzvald}
\affiliation{Department of Particle Physics and Astrophysics, Weizmann Institute of Science, Rehovot 76100, Israel}
% 
% Operation Team
% 7
\author{Sang-Mok Cha}
\affiliation{Korea Astronomy and Space Science Institute, Daejeon 34055, Republic of Korea}
\affiliation{School of Space Research, Kyung Hee University, Yongin, Kyeonggi 17104, Republic of Korea}
% 8 
\author{Dong-Jin Kim}
\affiliation{Korea Astronomy and Space Science Institute, Daejeon 34055, Republic of Korea}
% 9
\author{Hyoun-Woo Kim} % ONLY for events with years <= 2019
\affiliation{Korea Astronomy and Space Science Institute, Daejeon 34055, Republic of Korea}
% 10
\author{Seung-Lee Kim} 
\affiliation{Korea Astronomy and Space Science Institute, Daejeon 34055, Republic of Korea}
% 11
\author{Dong-Joo Lee}
\affiliation{Korea Astronomy and Space Science Institute, Daejeon 34055, Republic of Korea}
% 12
\author{Yongseok Lee}
\affiliation{Korea Astronomy and Space Science Institute, Daejeon 34055, Republic of Korea}
\affiliation{School of Space Research, Kyung Hee University, Yongin, Kyeonggi 17104, Republic of Korea}
% 13
\author{Byeong-Gon Park}
\affiliation{Korea Astronomy and Space Science Institute, Daejeon 34055, Republic of Korea}
% 14
\author{Richard W. Pogge}
\affiliation{Department of Astronomy, The Ohio State University, 140 W. 18th Ave., Columbus, OH 43210, USA}
\collaboration{14}{(The KMTNet Collaboration),}
% OGLE -------------------------------
% 1
\author{Przemek Mr{\'o}z}
\affiliation{Astronomical Observatory, University of Warsaw, Al. Ujazdowskie 4, 00-478 Warszawa, Poland}
% 2
\author{Micha{\l} K. Szyma{\'n}ski}
\affiliation{Astronomical Observatory, University of Warsaw, Al. Ujazdowskie 4, 00-478 Warszawa, Poland}
% 3
\author{Jan Skowron}
\affiliation{Astronomical Observatory, University of Warsaw, Al. Ujazdowskie 4, 00-478 Warszawa, Poland}
% 4
\author{Rados{\l}aw Poleski}
\affiliation{Astronomical Observatory, University of Warsaw, Al. Ujazdowskie 4, 00-478 Warszawa, Poland}
% 5
\author{Igor~Soszy{\'n}ski}
\affiliation{Astronomical Observatory, University of Warsaw, Al. Ujazdowskie 4, 00-478 Warszawa, Poland}
% 6
\author{Pawe{\l} Pietrukowicz}
\affiliation{Astronomical Observatory, University of Warsaw, Al. Ujazdowskie 4, 00-478 Warszawa, Poland}
% 7
\author{Szymon Koz{\l}owski}
\affiliation{Astronomical Observatory, University of Warsaw, Al. Ujazdowskie 4, 00-478 Warszawa, Poland}
% 8
\author{Krzysztof A. Rybicki}
\affiliation{Astronomical Observatory, University of Warsaw, Al. Ujazdowskie 4, 00-478 Warszawa, Poland}
\affiliation{Department of Particle Physics and Astrophysics, Weizmann Institute of Science, Rehovot 76100, Israel}
% 9
\author{Patryk Iwanek}
\affiliation{Astronomical Observatory, University of Warsaw, Al. Ujazdowskie 4, 00-478 Warszawa, Poland}
% 10
\author{Krzysztof Ulaczyk}
\affiliation{Department of Physics, University of Warwick, Gibbet Hill Road, Coventry, CV4 7AL, UK}
% 11
\author{Marcin Wrona}
\affiliation{Astronomical Observatory, University of Warsaw, Al. Ujazdowskie 4, 00-478 Warszawa, Poland}
% 12
\author{Mariusz Gromadzki}
\affiliation{Astronomical Observatory, University of Warsaw, Al. Ujazdowskie 4, 00-478 Warszawa, Poland}
\collaboration{13}{(The OGLE Collaboration)}
% MOA --------------------------------
% 1
\author{Fumio Abe}
\affiliation{Institute for Space-Earth Environmental Research, Nagoya University, Nagoya 464-8601, Japan}
% 2
\author{Ken Bando}
\affiliation{Department of Earth and Space Science, Graduate School of Science, Osaka University, Toyonaka, Osaka 560-0043, Japan}
% 3
\author{Richard Barry}
\affiliation{Code 667, NASA Goddard Space Flight Center, Greenbelt, MD 20771, USA}
% 4
\author{David P. Bennett}
\affiliation{Code 667, NASA Goddard Space Flight Center, Greenbelt, MD 20771, USA}
\affiliation{Department of Astronomy, University of Maryland, College Park, MD 20742, USA}
% 5
\author{Aparna Bhattacharya}
\affiliation{Code 667, NASA Goddard Space Flight Center, Greenbelt, MD 20771, USA}
\affiliation{Department of Astronomy, University of Maryland, College Park, MD 20742, USA}
% 6
\author{Ian A. Bond}
\affiliation{Institute of Natural and Mathematical Sciences, Massey University, Auckland 0745, New Zealand}
% 7
\author{Hirosane Fujii}
\affiliation{Institute for Space-Earth Environmental Research, Nagoya University, Nagoya 464-8601, Japan}
% 8
\author{Akihiko Fukui}
\affiliation{Department of Earth and Planetary Science, Graduate School of Science, The University of Tokyo, 7-3-1 Hongo, Bunkyo-ku, Tokyo 113-0033, Japan}
\affiliation{Instituto de Astrof\'isica de Canarias, V\'ia L\'actea s/n, E-38205 La Laguna, Tenerife, Spain}
% 9
\author{Ryusei Hamada}
\affiliation{Department of Earth and Space Science, Graduate School of Science, Osaka University, Toyonaka, Osaka 560-0043, Japan}
% 10
\author{Shunya Hamada}
\affiliation{Department of Earth and Space Science, Graduate School of Science, Osaka University, Toyonaka, Osaka 560-0043, Japan}
% 11
\author{Naoto Hamasaki}
\affiliation{Department of Earth and Space Science, Graduate School of Science, Osaka University, Toyonaka, Osaka 560-0043, Japan}
% 12
\author{Yuki Hirao}
\affiliation{Institute of Astronomy, Graduate School of Science, The University of Tokyo, 2-21-1 Osawa, Mitaka, Tokyo 181-0015, Japan}
% 13
\author{Stela Ishitani Silva}
\affiliation{Department of Physics, The Catholic University of America, Washington, DC 20064, USA}
\affiliation{Code 667, NASA Goddard Space Flight Center, Greenbelt, MD 20771, USA}
% 14
\author{Yoshitaka Itow}
\affiliation{Institute for Space-Earth Environmental Research, Nagoya University, Nagoya 464-8601, Japan}
% 15
\author{Rintaro Kirikawa}
\affiliation{Department of Earth and Space Science, Graduate School of Science, Osaka University, Toyonaka, Osaka 560-0043, Japan}
% 16
\author{Naoki Koshimoto}
\affiliation{Department of Earth and Space Science, Graduate School of Science, Osaka University, Toyonaka, Osaka 560-0043, Japan}
% 17
\author{Yutaka Matsubara}
\affiliation{Institute for Space-Earth Environmental Research, Nagoya University, Nagoya 464-8601, Japan}
% 18
\author{Shota Miyazaki}
\affiliation{Institute of Space and Astronautical Science, Japan Aerospace Exploration Agency, 3-1-1 Yoshinodai, Chuo, Sagamihara, Kanagawa 252-5210, Japan}
% 19
\author{Yasushi Muraki}
\affiliation{Institute for Space-Earth Environmental Research, Nagoya University, Nagoya 464-8601, Japan}
% 20
\author{Tutumi NAGAI}
\affiliation{Department of Earth and Space Science, Graduate School of Science, Osaka University, Toyonaka, Osaka 560-0043, Japan}
% 21
\author{Kansuke NUNOTA}
\affiliation{Department of Earth and Space Science, Graduate School of Science, Osaka University, Toyonaka, Osaka 560-0043, Japan}
% 22
\author{Greg Olmschenk}
\affiliation{Code 667, NASA Goddard Space Flight Center, Greenbelt, MD 20771, USA}
% 23
\author{Cl\'ement Ranc}
\affiliation{Sorbonne Universit\'e, CNRS, UMR 7095, Institut d'Astrophysique de Paris, 98 bis bd Arago, 75014 Paris, France}
% 24
\author{Nicholas J. Rattenbury}
\affiliation{Department of Physics, University of Auckland, Private Bag 92019, Auckland, New Zealand}
% 25
\author{Yuki Satoh}
\affiliation{Department of Earth and Space Science, Graduate School of Science, Osaka University, Toyonaka, Osaka 560-0043, Japan}
% 26
\author{Daisuke Suzuki}
\affiliation{Department of Earth and Space Science, Graduate School of Science, Osaka University, Toyonaka, Osaka 560-0043, Japan}
% 27
\author{Mio Tomoyoshi}
\affiliation{Department of Earth and Space Science, Graduate School of Science, Osaka University, Toyonaka, Osaka 560-0043, Japan}
% 28
\author{Paul . J. Tristram}
\affiliation{University of Canterbury Mt.\ John Observatory, P.O. Box 56, Lake Tekapo 8770, New Zealand}
% 29
\author{Aikaterini Vandorou}
\affiliation{Code 667, NASA Goddard Space Flight Center, Greenbelt, MD 20771, USA}
\affiliation{Department of Astronomy, University of Maryland, College Park, MD 20742, USA}
% 30
\author{Hibiki Yama}
\affiliation{Department of Earth and Space Science, Graduate School of Science, Osaka University, Toyonaka, Osaka 560-0043, Japan}
% 31
\author{Kansuke Yamashita}
\affiliation{Department of Earth and Space Science, Graduate School of Science, Osaka University, Toyonaka, Osaka 560-0043, Japan}
\collaboration{32}{(the MOA Collaboration)}
% ------------------------------------------------------------------------------------------------------------------------

\begin{abstract}
Following \citet{shin23b}, which is a part of the Systematic KMTNet Planetary Anomaly Search series (i.e., a search for planets in the 2016 KMTNet prime fields), we conduct a systematic search of the 2016 KMTNet sub-prime fields using a semi-machine-based algorithm to identify hidden anomalous events missed by the conventional by-eye search. We find four new planets and seven planet candidates that were buried in the KMTNet archive. The new planets are \fifteenninetyeight Lb, \eighteendoubleO Lb, \fivetwentysix Lb, and \twentythreetwentyone Lb, which show typical properties of microlensing planets, i.e., giant planets orbit M dwarf host stars beyond their snow lines. For the planet candidates, we find planet/binary or 2L1S/1L2S degeneracies, which are an obstacle to firmly claiming planet detections. By combining the results of \citet{shin23b} and this work, we find a total of nine hidden planets, which is about half the number of planets discovered by eye in 2016. With this work, we have met the goal of the systematic search series for 2016, which is to build a complete microlensing planet sample. We also show that our systematic searches significantly contribute to completing the planet sample, especially for planet/host mass ratios smaller than $10^{-3}$, which were incomplete in previous by-eye searches of the KMTNet archive.
\end{abstract}

\section{Introduction}
Since 2016, the Korea Microlensing Telescope Network \citep[KMTNet;][]{kim16} has operated a microlensing survey to detect exoplanets using their near-continuous observations toward the Galactic bulge. As of $2023$, the KMTNet has contributed to the discovery/characterization of more than $135$ microlensing planets\footnote{We count the discovered microlensing planets using the NASA Exoplanet Archive (\url{https://exoplanetarchive.ipac.caltech.edu/}) as of October 2023.}. Initially, the planetary events were identified by a traditional method, i.e., ``by-eye" search.

The human dependence of that method, which relies on the experience or insight of operators, is difficult to quantify and there may exist missing or hidden planets. Thus, we conduct a series of works called ``Systematic KMTNet Planetary Anomaly Search" to find {\it hidden} planets in the KMTNet data archive in order to build a complete microlensing planet sample. The complete sample can be used for statistical studies such as the planet frequency and mass-ratio distribution of planetary systems in our Galaxy.   

To systematically search anomalous events in the KMTNet data archive, we use a semi-machine-based algorithm called AnomalyFinder \citep[AF;][]{zang21,zang22} instead of the by-eye search. The AF search is separately conducted for each year and cadence. The nominal cadences of the KMTNet observations have two categories, which are high cadence ($\Gamma = 2.0-4.0 {\rm hr^{-1}}$ for prime fields) and low cadence ($\Gamma = 0.2-1.0 {\rm hr^{-1}}$ for sub-prime fields). The detailed information of the KMTNet fields is described in \citet{kim18}.

Based on the AF searches, we conducted detailed light curve analyses for the identified anomalous events. The parts of this AF series have been published or submitted. Indeed, from the systematic search, we can find hidden planets that are missing from the by-eye search. \citet{shin23b} reported 5 planets, which were newly found in the 2016 prime fields. \citet{ryu23} is submitted to report 3 new planets found in the 2017 prime fields\footnote{Among the three planetary events, two events were newly identified by the AF. While one event was previously identified by eye. However, this event was not published due to technical complications.}. \citet{gould22, wang22, hwang22} reported a total of 12 new planets, which were discovered in the 2018 prime fields. \citet{jung22} reported 6 new planets found in the 2018 sub-prime fields. \citet{zang21, zang22, hwang22} reported a total of 7 new planets discovered in the 2019 prime fields. \citet{jung23} reported 5 new planets found in the 2019 sub-prime fields. Lastly, \citet{zang23} present 7 new planets having $q < 10^{-4}$, which were identified by the AF in the KMTNet data archive observed from 2016 to 2019. Although our systematic search works are not complete, yet, we have found a total of 45 hidden planets in the KMTNet archive, which amounts to about $33\%$ of the total microlensing planets discovered from 2016 to 2022. 

Following the work of \citet{shin23b}, we conduct the AF search for 2016 sub-prime fields to find hidden planetary systems. The AF identifies a total of $113$ anomalous events in the fields, including recovery of all previously published planetary events identified by eye. Among them, we find that $91$ events were caused by binary lens systems (i.e., $q > 0.06$) from the preliminary light curve analyses using the KMTNet pipeline data. For the remaining $22$ events, we conduct detailed light curve analyses using the re-reduced data sets with the best quality (see Section \ref{sec:obs}). The detailed analyses reveal that $11$ events do not have possible planetary solutions (i.e., $q < 0.03$; see Appendix \ref{sec:appendix_binaries}). Finally, we find $4$ new planetary events and $7$ planet candidates on the $2016$ sub-prime fields. The new planets are \fifteenninetyeight Lb, \eighteendoubleO Lb, \fivetwentysix Lb, and \twentythreetwentyone Lb. We present the detailed light curve analyses for these planetary events in Section \ref{sec:LC_analysis}. In this section, we also present the analyses of the planet candidates to show the possibility of planet detection. In Section \ref{sec:CMDs} and \ref{sec:lens_properties}, we present the analyses of color-magnitude diagrams and lens properties of each planetary system, respectively. Lastly, we summarize our findings in Section \ref{sec:summary}.

\section{Observations} \label{sec:obs}
Although the AF identified anomalous events based on the KMTNet data archive, these events may also have been independently observed or discovered by other microlensing surveys. Thus, we gather all available data sets for each event. In Tables \ref{table:obs_planet}, we list anomalous events that have at least one solution with $q < 0.06$ from the preliminary analyses along with their observational information. Note that, following the standard convention, we designate them according to the survey that first announced the event. 

The KMTNet data sets were obtained from three identical $1.6$-m telescopes equipped with $4$-square degree wide field cameras, which are located at three sites in the southern hemisphere, i.e., the Cerro Tololo Inter-American Observatory in Chile (KMTC), South African Astronomical Observatory in South Africa (KMTS), and Siding Spring Observatory in Australia (KMTA). Note that, in the figures, the two-digit number after the site acronym indicates the field number of the KMTNet survey. These sites cover well-separated time zones to achieve near-continuous observations. The KMTNet observations are initially reduced using their pySIS pipeline \citep{albrow09}, which adopts the difference image analysis method \citep[DIA;][]{tomaney96, alard98}. For KMTC and KMTS observations, the KMTNet survey regularly takes one $V$-band observation for every $10$-th and $20$-th $I$-band observations, respectively (Johnson-Cousins $BVRI$ filter system). The pipeline data are available at the KMTNet alert System \citep[\url{https://kmtnet.kasi.re.kr/~ulens/}]{kim18}. 

Note that we manually re-reduced the KMTNet data sets for each preliminary planet candidate listed in Table \ref{table:obs_planet} using the updated pySIS package described in \citet{yang23}. We conduct light curve analyses based on these tender-loving-care (TLC) reductions, which have checked the anomalous data points with the best quality.

The OGLE data sets were obtained from a $1.6$-m Warsaw telescope equipped with a $1.4$ square degree field camera, which is located at Las Campanas Observatory in Chile. For the OGLE observations, it mainly takes $I$-band observations and periodically takes $V$-band observations. The OGLE observations are reduced by their own DIA pipeline \citep{wozniak00}. The data are available on the OGLE Early Warning System \citep[\url{http://ogle.astrouw.edu.pl/ogle4/ews/ews.html}]{udalski94}.

The MOA data sets were obtained from a $1.8$-m telescope located at Mt. John University Observatory in New Zealand. The observations were made in the MOA-Red band (hereafter, referred to as $R$ band), which has wavelength ranges of $609-1109$ nm and transmission ranges of $0.0-0.978$ (i.e., a rough sum of the Cousins $R$ and $I$ bands). The MOA observations were reduced by their DIA pipeline \citep{bond01}, which are available on the MOA alert system (\url{http://www.massey.ac.nz/~iabond/moa/alerts/}).

\section{Light curve Analysis} \label{sec:LC_analysis}

\subsection{Basics of the Light Curve Analysis}
We conduct light curve analysis following the procedures described in \citet{shin23b}, which describes the systematic KMTNet planetary anomaly search for $2016$ prime-field events. To avoid redundant descriptions for analysis procedures, we do not present the details here. However, in Table \ref{table:definitions}, we present definitions of acronyms and model parameters to describe the analysis results in the following sections. We note that we test the APRX effect if the event has a relatively long timescale, which is defined as larger than $t_{\rm E} > 15.0$ days. Once we detect the APRX effect, we also test the OBT effect and xallarap effect to confirm the robustness of the APRX detection. Because the OBT can affect the APRX measurement and its uncertainty and the xallarap can mimic the APRX effect. Lastly, if we find a planetary solution(s) from bump-shaped anomalies on the light curve, we test the 2L1S/1L2S degeneracy \citep{gaudi98} to confirm planet detection.

\subsection{Planetary Events}
We find four events caused by planetary lens systems that satisfy our criteria to claim planet detection. For clarity, we summarize our criteria to claim planet detection as follows:
\begin{itemize}
    \item[a)] The mass ratio of the best-fit planetary solution must be smaller than $0.03$ (i.e., $q < 0.03$).
    \item[b)] Competing binary-lens solutions can be resolved by $\Delta\chi^{2} > 10.0$.
    \item[c)] If the 2L1S/1L2S degeneracy exists, 1L2S can be resolved by $\Delta\chi^{2} > 15.0$.
\end{itemize}
We present the details of light curve analysis for each planetary event in the following sections.

\subsubsection{\fifteenninetyeight} % Planet : KMT-2016-BLG-0696 | OGLE-2016-BLG-1598(*) | MOA-2016-BLG-521
\begin{figure}[htb!]
\epsscale{1.00}
\plotone{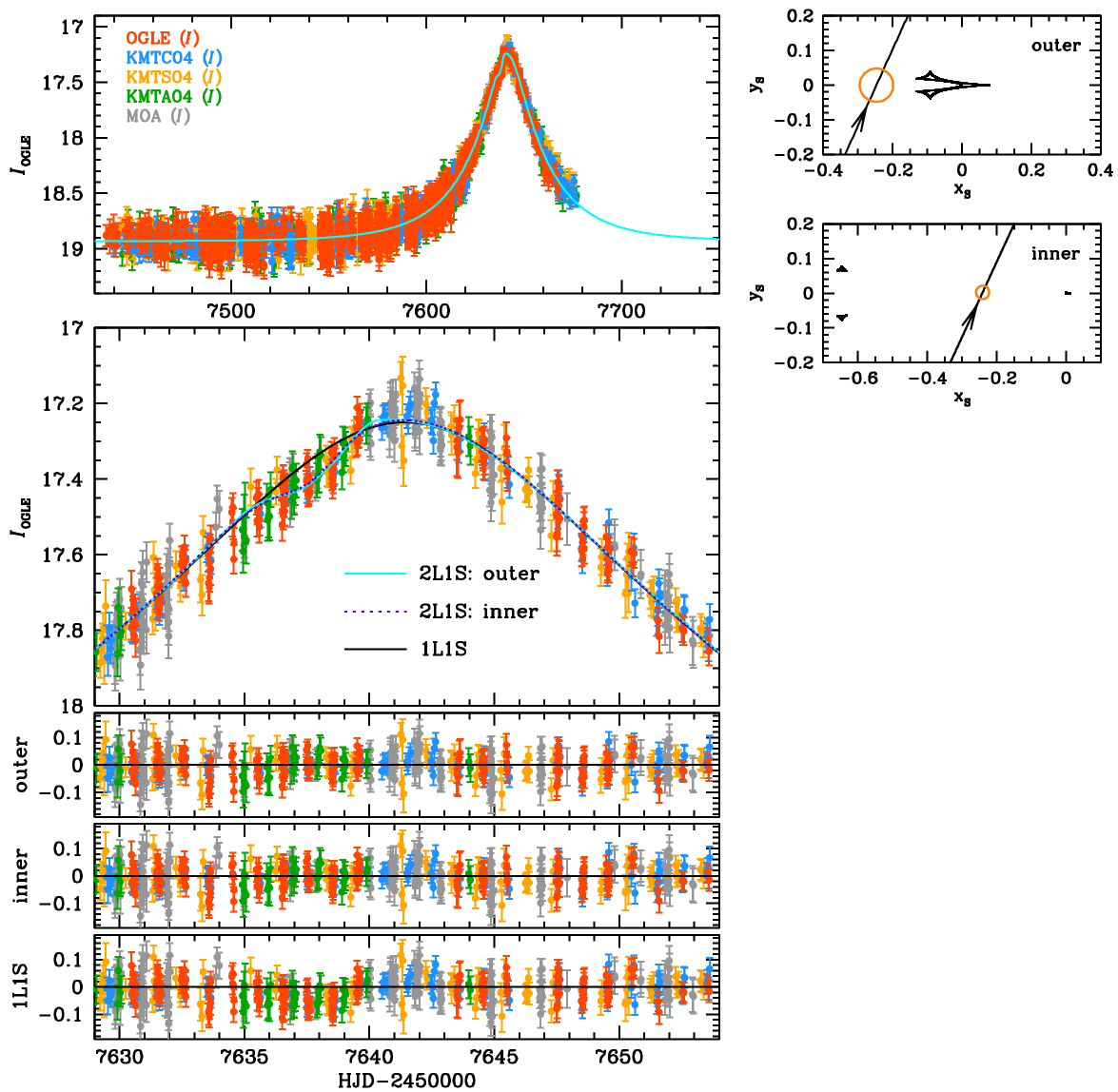}
\caption{Light curve of \fifteenninetyeight\ with 2L1S and 1L1S models. We also present caustic geometries 
of the 2L1S models. 
\label{fig:lc_0696}}
\end{figure}

As shown in Figure \ref{fig:lc_0696}, the light curve of \fifteenninetyeight\ (which we identified as KMT-2016-BLG-0696) exhibits a shallow--dip anomaly near the peak (i.e., HJD$^{\prime} = 7636.0 \sim 7639.0$), which shows clear residuals from the 1L1S model (i.e., $\Delta\chi^{2}_{\rm 1L1S - 2L1S} = 171.9$). The anomaly can be explained by two 2L1S models caused by the inner/outer degeneracy. Although the degenerate models cannot be resolved (i.e., $\Delta\chi^{2} = 8.5$), both cases indicate that the lens system is a planetary system (i.e., $q < 0.03$) as presented in Table \ref{table:model_0696}. Thus, we conclude that \fifteenninetyeight\ was caused by a planetary lens system. Indeed, the heuristic analysis ($t_{\rm anom} = 7637.5$, $\tau_{\rm anom} = -0.1045$, and $u_{\rm anom} = 0.2436$) predicts $s^{\dagger}_{-} = 0.886$, $s^{\dagger}_{+} = 1.129$, and $q \sim 6.4\times10^{-4}$. The predicted $q$ is consistent with the empirical $q = 6.44\times10^{-4}$ value. Also, the empirical $s^{\dagger} = \sqrt{s_{\rm inner}s_{\rm outer}} = 0.833$ is similar to the predicted $s_{-}^{\dagger}$ value.

\begin{figure}[htb!]
\epsscale{1.00}
\plotone{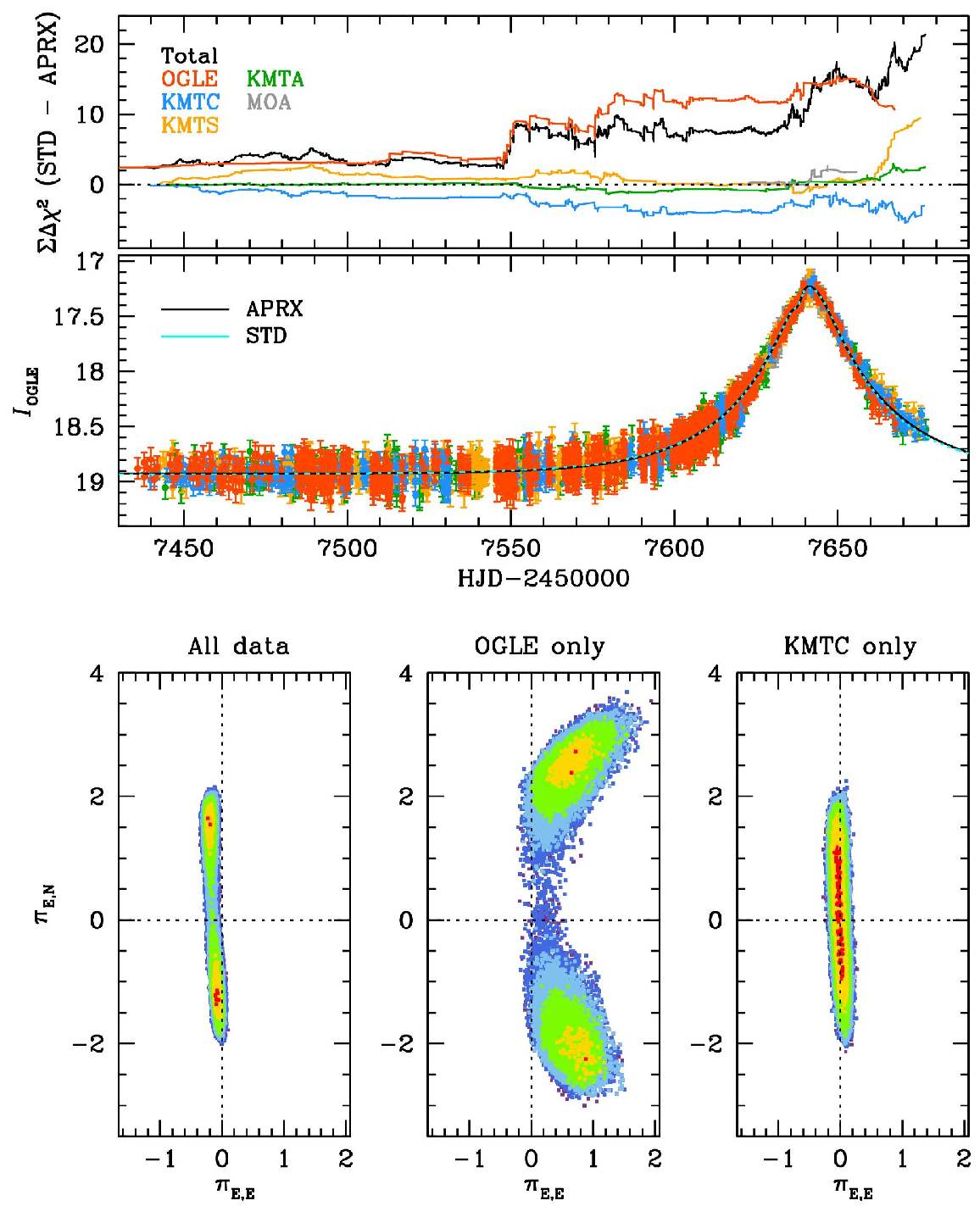}
\caption{APRX test of \fifteenninetyeight. The upper two panels show the cumulative $\Delta\chi^{2}$ plot 
between the APRX and STD models with the light curve. The lower three panels show APRX contours obtained 
using all data (left), OGLE only (middle), and KMTC only (right), respectively.
\label{fig:APRX_0696}}
\end{figure}

Because of the relatively long timescale ($t_{\rm E} \sim 38$ days) for all cases, we test the annual microlens parallax (APRX) effect. As shown in Figure \ref{fig:APRX_0696}, we find the $\chi^{2}$ improves by $\sim 20.7$ when we consider the APRX effect, which mostly comes from the OGLE data. However, the improvement of the OGLE data is inconsistent with the KMTNet and MOA data. Moreover, there is no improvement in the case of the KMTC data although the KMTC data have similar coverage to the OGLE data. Thus, we separately conduct APRX modeling using KMTC and OGLE only. We find that the OGLE--only case favors too large APRX values (i.e., $|\pivec| > 2.82$), which are unreliable. In contrast, the KMTC--only case shows that the APRX values are consistent with a non-detection (i.e., $(\pieE, \pieN) \sim (0.0, 0.0)$ within $1\sigma$ level). The inconsistency between OGLE and KMTNet data of both the $\chi^{2}$ improvements and the APRX measurements indicates that the APRX effect of this event is unreliable. We test again the APRX effect using re-reduced OGLE data. Even though we use the best--quality data sets, we have the same results from the test. Hence, we conclude that the STD models should be the fiducial solutions for this event.

\subsection{\eighteendoubleO} % Planet : KMT-2016-BLG-0781 | OGLE-2016-BLG-1800(*) | MOA-2016-BLG-581
\begin{figure}[htb!]
\epsscale{1.00}
\plotone{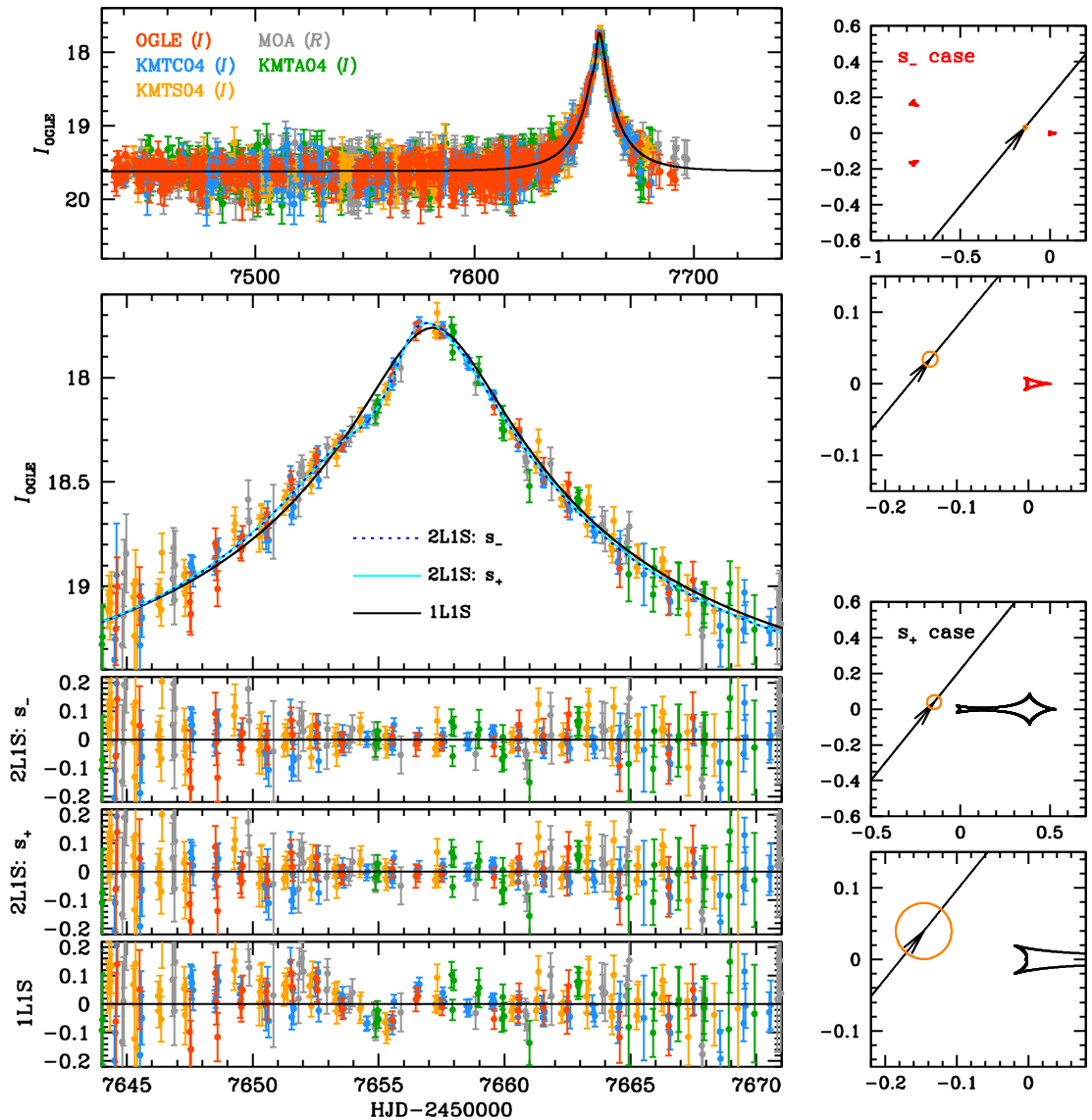}
\caption{Light curve of \eighteendoubleO\ with both 2L1S solutions compared to the 1L1S models.
\label{fig:lc_0781}}
\end{figure}

In Figure \ref{fig:lc_0781}, we present the light curve of \eighteendoubleO\ (which we identified as KMT-2016-BLG-0781), which shows deviations (HJD$^{\prime} = 7651. - 7657.0$) from the 1L1S model. The anomaly can be explained by the 2L1S models that fit better by $\Delta\chi^{2} = 196.33$ compared to the 1L1S fits. In Table \ref{table:model_0781}, we present the model parameters of the 2L1S solutions. Indeed, the heuristic analysis predicts $s_{-}^{\dagger} = 0.918$ and $s_{+}^{\dagger} = 1.090$ from $\tau_{\rm anom} = -0.100$, $u_{\rm anom} = 0.172$, which is similar to the value of $s^{\dagger} = \sqrt{s_{-}s_{+}} = 0.911$ from the models.

We find that the $s_{\pm}$ cases of the 2L1S solutions cannot be resolved ($\Delta\chi^{2} = 0.92$). However, the mass ratios of both solutions indicate that the lens system consists of a planet and a host star. Thus, we conclude that \eighteendoubleO\ was caused by the planetary lens system.

Because the timescales of both cases are relatively long (i.e., $t_{\rm E} \sim 20$ days), we test the APRX effect for this event. However, we find the $\chi^{2}$ improvement is negligible, only $\Delta\chi^{2} = 0.74$. Thus, we treat the STD cases as the fiducial solutions for this event. Also, for both cases, the $\rho_{\ast}$ is not measured as expected from the non-caustic-crossing geometries (see geometries in Figure \ref{fig:lc_0781}).

\subsubsection{\fivetwentysix} % Planet : KMT-2016-BLG-1611 | OGLE-2016-BLG-1705 | MOA-2016-BLG-526(*)
\begin{figure}[htb!]
\epsscale{1.00}
\plotone{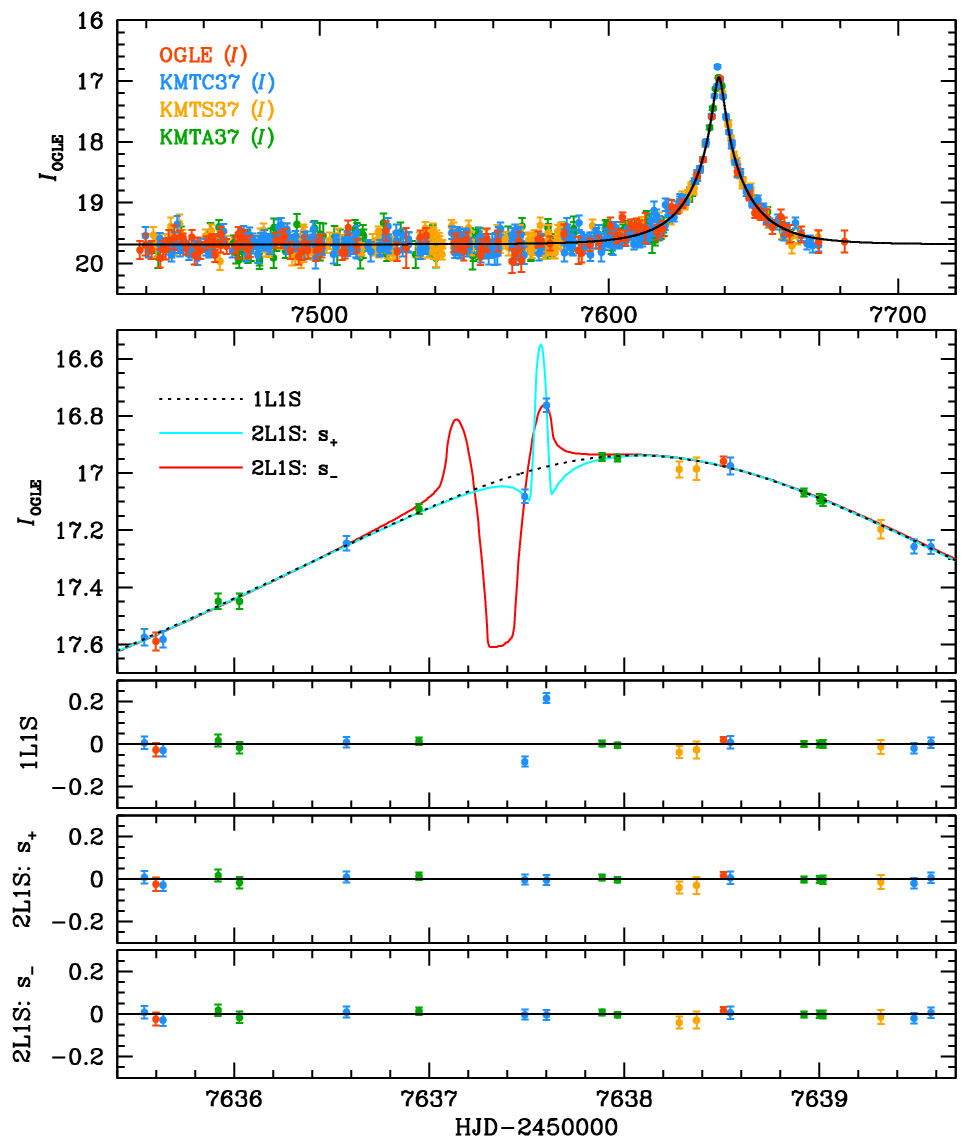}
\caption{Light curve of \fivetwentysix\ with 2L1S solutions compared to the 1L1S models.
\label{fig:lc_1611_01}}
\end{figure}

As shown in Figure \ref{fig:lc_1611_01}, in the light curve of \fivetwentysix\ (which we identified as KMT-2016-BLG-1611), two KMTC points near the peak exhibit an anomaly from the 1L1S model\footnote{We note that MOA data did not cover the anomaly part although the MOA firstly announced this event. Also, the data have systematics that might be caused by the faintness of the source or bad weather conditions. Thus, we do not include the MOA data in the analysis.}. Based on the TLC reductions, we investigate these points to check whether or not the anomaly is reliable. We find that the anomalous points are robust. Thus, we conduct the 2L1S modeling to describe the anomaly. We find that the 2L1S models can perfectly explain the anomaly, which shows better fits by $\Delta\chi^{2} \sim 84$ compared to the 1L1S model. 

Because of the sparse coverage, we find that there exist several degenerate solutions as presented in Table \ref{table:model_1611}. Indeed, we predict $s_{-}^{\dagger} = 0.954$, $s_{+}^{\dagger} = 1.048$, and $q \sim 2.9\times10^{-4}$ from the heuristic analysis ($\tau_{\rm anom} = -0.026$ and $u_{\rm anom} = 0.094$). The $s_{-}^{\dagger}$ is consistent with the $s^{\dagger} = \sqrt{s_{-}(A)s_{-}(C)} = 0.954$ and the $s$ value of the $s_{-}$ (B) case. The $s_{+}^{\dagger}$ is also consistent with $s^{\dagger} = \sqrt{s_{+}(A)s_{+}(B)} = 1.049$. The predicted $q$ is similar to empirical $q$ values of the $s_{-}$ cases by a factor of $\sim 2$. 

\begin{figure}[htb!]
\epsscale{1.00}
\plotone{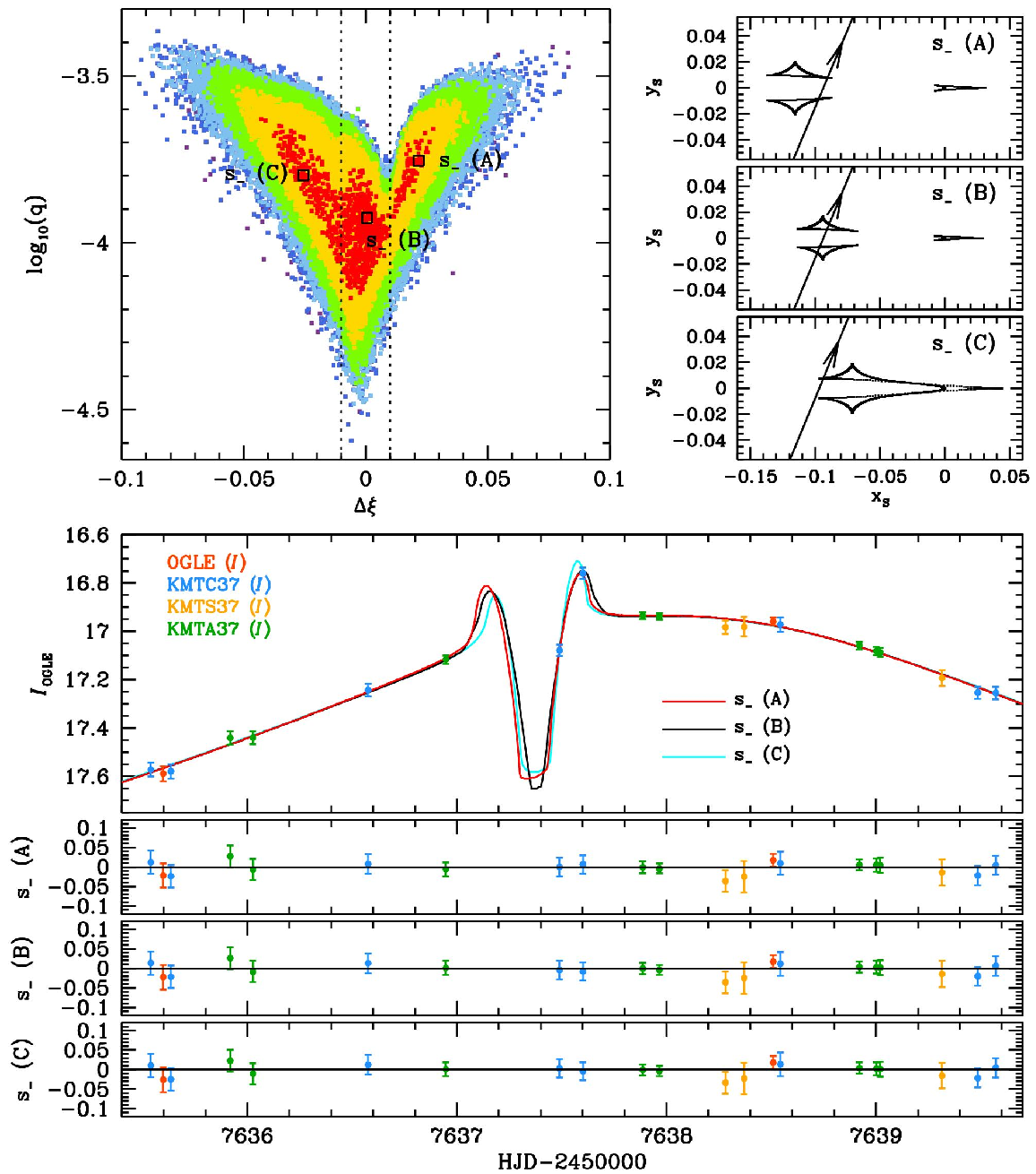}
\caption{Light curve of \fivetwentysix\ with the family of degenerate 2L1S $s_{-}$ models with the 
$\Delta\xi-\log_{10}(q)$ space. In the $\Delta\xi-\log_{10}(q)$ space (upper-left panel), each color 
represents $\Delta\chi^{2} \le n^{2}$ from the best-fit $\chi^{2}$ where $n = 1$ (red), $2$ (yellow), 
$3$ (green), $4$ (light blue), $5$ (blue), and $6$ (purple), respectively.
\label{fig:lc_1611_02}}
\end{figure}

\begin{figure}[htb!]
\epsscale{1.00}
\plotone{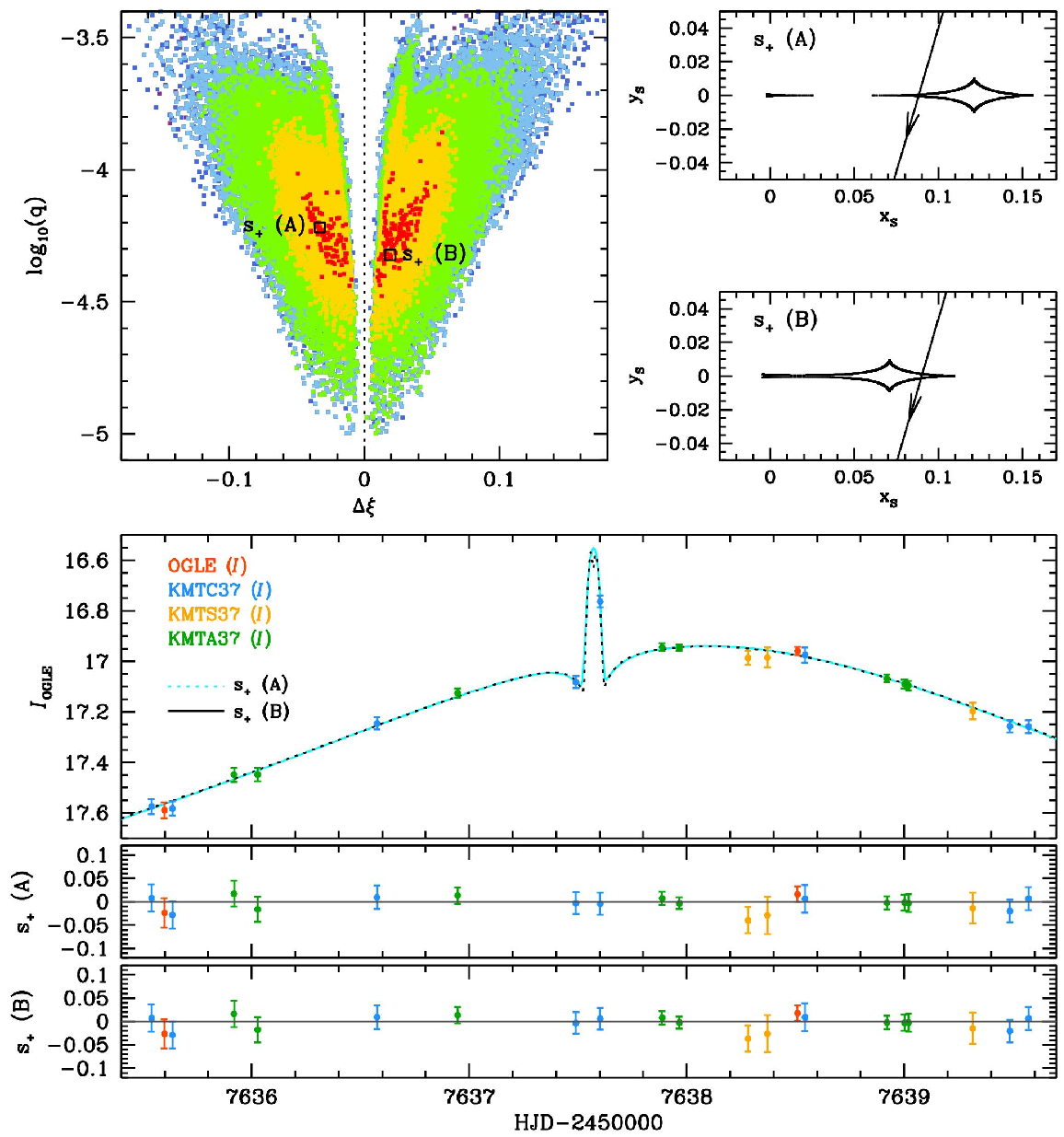}
\caption{Light curve of \fivetwentysix\ with the degenerate 2L1S $s_{+}$ models and 
the $\Delta\xi-\log_{10}(q)$ space.The color scheme of the $\Delta\xi-\log_{10}(q)$ space is identical 
to that of Figure \ref{fig:lc_1611_02}.
\label{fig:lc_1611_03}}
\end{figure}

For the $s_{-}$ case, we find several degenerate solutions within $\Delta\chi^{2} < 1.0$. These solutions show three categories of geometries as shown in Figure \ref{fig:lc_1611_02}. The A, B, and C families are produced by the different source trajectories, which travel over the inner, intermediate, and outer parts of the caustics, respectively. Indeed, this kind of degeneracy was introduced in the analysis of OGLE-2017-BLG-0173Lb \citep{hwang18}. Thus, we adopt the $\Delta\xi$ ($\equiv u_{0}\csc{\alpha} - [s -s^{-1}]$) parameter described in \citet{hwang18} to separate and extract each family (see dotted lines in the upper-left panel of Figure \ref{fig:lc_1611_02}). We present the best-fit solution for each family as a representative. For the $s_{+}$ case, we find two solutions caused by the inner/outer degeneracy, which cannot be distinguished (i.e., $\Delta\chi^{2} = 0.2$). In Figure \ref{fig:lc_1611_03}, we present the light curves of $s_{+}$ solutions with their geometries. For consistency, we also present $\Delta\xi - \log_{10}(q)$ space to show the locations of each solution, which are clearly divided into two categories. Although there exist several degenerate solutions with $\Delta\chi^{2} \lesssim 1.0$, all solutions have mass ratios less than $0.03$. Thus, we conclude that this event was caused by a planetary lens system.

Because of the relatively long timescale ($t_{\rm E} \sim 20$ days) for all solutions, we test the APRX effect for this event. However, we find a negligible $\chi^{2}$ improvement of $\Delta\chi^{2} = 1.55$ compared to the best-fit of STD solution and no meaningful constraints on $\pivec$. Thus, we treat the STD models as our fiducial solutions. Note that, because of the sparse coverage, we cannot measure the $\rho_{\ast}$ for all STD cases even though some cases show caustic-crossing features.

Lastly, the $s_{+}$ solutions exhibit a bump-like anomaly, which can yield a 2L1S/1L2S degeneracy. Thus we check whether or not the 1L2S model can explain the anomaly. We find that the 1L2S model cannot explain the KMTC point at HJD$^{\prime} = 7637.4903$, which shows a shallow dip relative to the 1L1S fits. Also, we find that the 1L2S interpretation is fine-tuned to describe the KMTC point at HJD$^{\prime} = 7637.6021$. That is, to fit this point, the 1L2S model has $q_{\rm flux} \sim \mathcal{O}(10^{-4})$, which is nonphysical. Thus, we conclude that there is no 2L1S/1L2S degeneracy for this event despite the fact that, due to the lack of covered data points, $\Delta\chi^{2} = 11.7$ between the 2L1S and 1L2S models is smaller than our formal threshold.

\subsubsection{\twentythreetwentyone} % Planet : KMT-2016-BLG-2321
\begin{figure}[htb!]
\epsscale{1.00}
\plotone{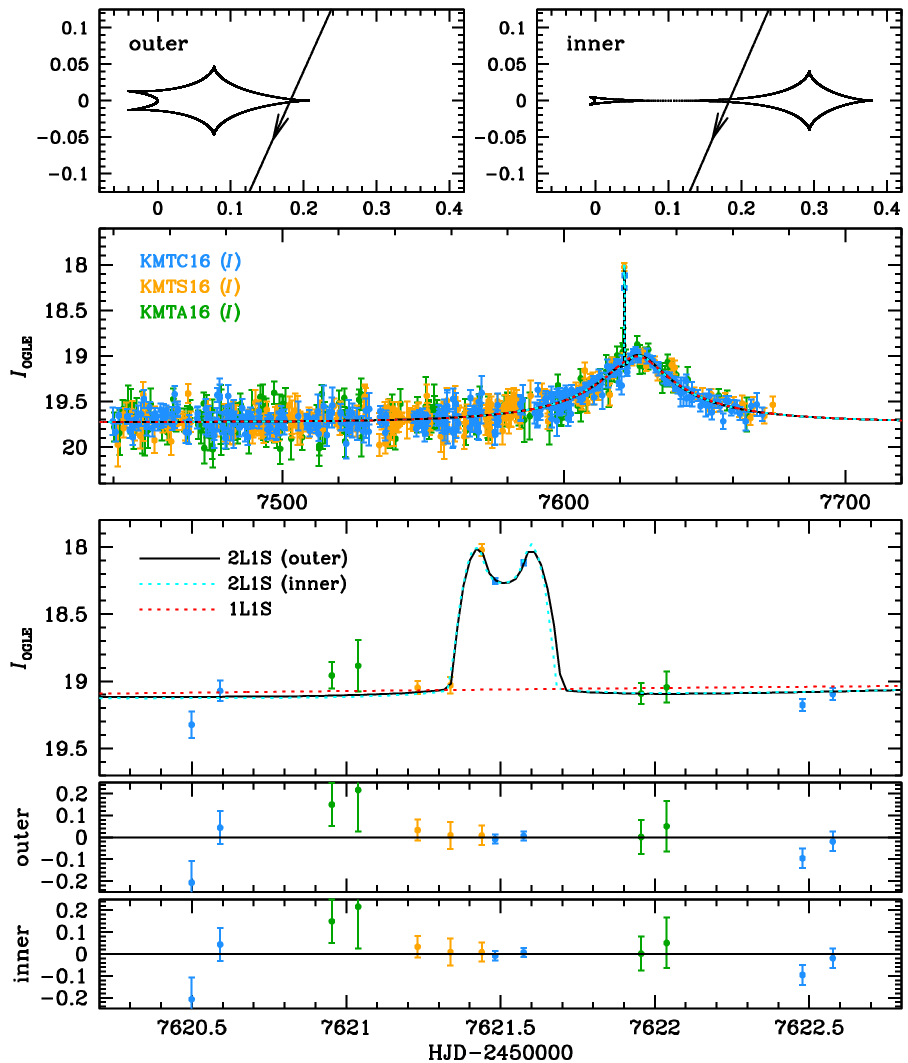}
\caption{Light curve of \twentythreetwentyone\ with the 2L1S solutions compared to the 1L1S models.
\label{fig:lc_2321}}
\end{figure}

As shown in Figure \ref{fig:lc_2321}, the light curve of \twentythreetwentyone\ exhibits an apparent anomaly at HJD$^{\prime} \sim 7621.5$ that has a short duration ($\sim 0.35$ days). We find that the anomaly can be explained by 2L1S models with caustic-crossing geometries. In Table \ref{table:model_2321}, we present the best-fit parameters of the 2L1S solutions. Indeed, we predict $s_{-}^{\dagger} = 0.912$ and $s_{+}^{\dagger} = 1.097$ from the heuristic analysis ($\tau_{\rm anom} = -0.0737$, $t_{\rm anom} = 0.1853$). The predicted $s_{+}^{\dagger}$ value corresponds with the empirical value of $s^{\dagger} = \sqrt{s_{\rm outer}s_{\rm inner}} = 1.096$. Although the 2L1S solutions caused by the inner/outer degeneracy cannot be distinguished ($\Delta\chi^{2} = 0.66$), the mass ratios of both solutions indicate that the event was caused by a planetary lens system, i.e., $q \sim \mathcal{O}(10^{-3})$.  

Because of the long timescales ($\sim 57$ days) for both solutions, we test the APRX effect. However, we find no $\chi^{2}$ improvement (the STD best-fit solution shows better fits than the APRX model by $\Delta\chi^{2} = 0.33$). Even though we additionally include the OBT effect (i.e., APRX+OBT model), we find a negligible $\chi^{2}$ improvement of $\Delta\chi^{2} = 1.28$ and no meaningful constraints on $\pivec$. Thus, we conclude that the higher-order effects are not available for this event. We note that, despite caustic-crossing features, the $\rho_{\ast}$ measurements are uncertain because the data coverage is not optimal.

Because of the caustic-crossing feature, we expect the 2L1S/1L2S degeneracy will not be an obstacle to claim planet detection. However, because the coverage is not optimal, we check the 1L2S model for confirmation. As expected, we find the 1L2S model is disfavored by $\Delta\chi^{2} = 132.56$, which cannot explain the caustic-crossing feature despite the non-optimal coverage.

\subsection{Planet Candidates}
We find $7$ planet candidates among the $11$ events, which are analyzed using the TLC data sets. These events have the possibility to be caused by a planetary lens system. However, these candidates cannot satisfy all our criteria to firmly claim planet detection. For example, there exist competing binary-lens solutions that cannot be resolved, or there is the 2L1S/1L2S degeneracy to prevent claiming the planet detection. Although we cannot firmly claim planet detection, there still remains the possibility that these events might be caused by a planetary system unless we have clear evidence against this conclusion. Hence, we report these planet candidates with the details of the light curve analyses for the record, in case there is an opportunity to conclusively reveal their nature in the future.

\subsubsection{\twelvefourtythree} % Planet Candidate : KMT-2016-BLG-1243 
\begin{figure}[htb!]
\epsscale{1.00}
\plotone{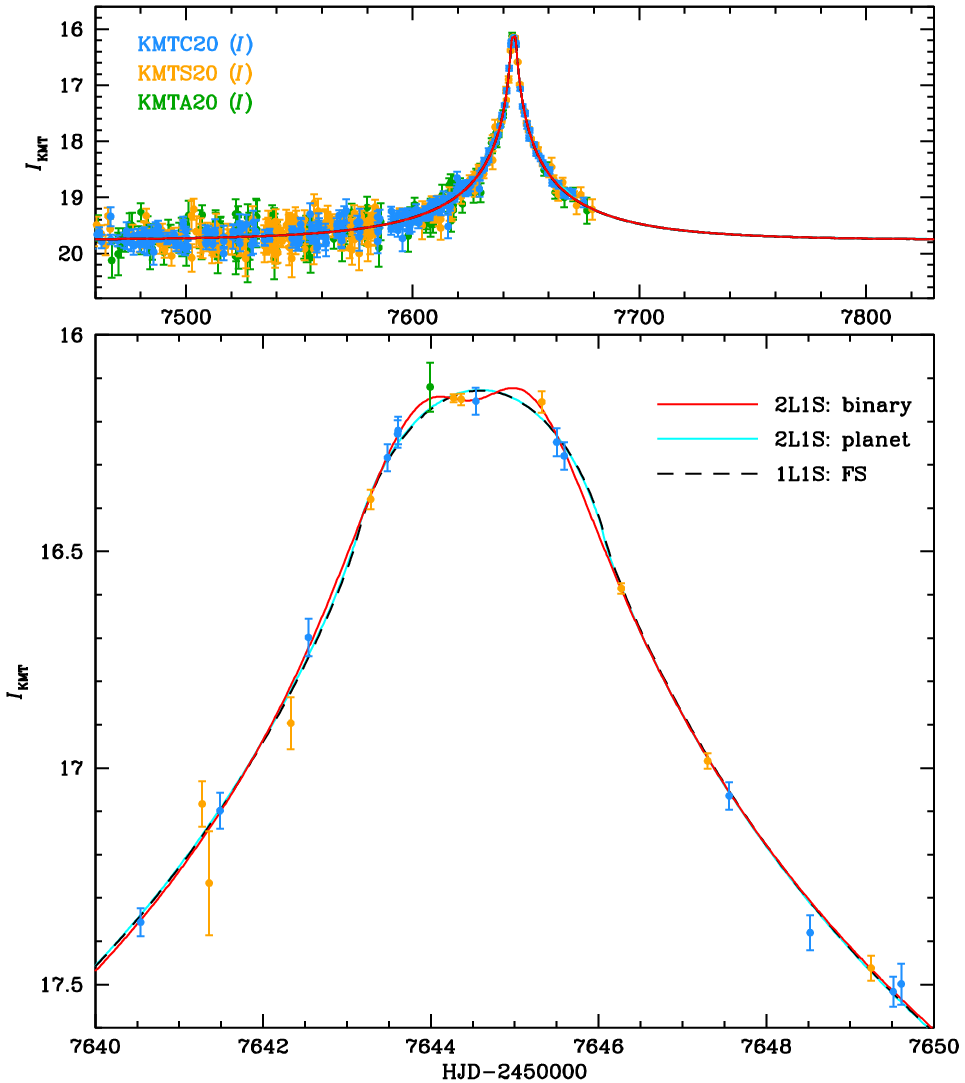}
\caption{Light curve of \twelvefourtythree\ with 2L1S binary and planet models compared to 
the 1L1S model. The 1L1S model includes the finite-source effect (FS).
\label{fig:lc_1243_01}}
\end{figure}

The light curve of \twelvefourtythree\ exhibits subtle deviations at the peak as shown in Figure \ref{fig:lc_1243_01}. The anomaly can be explained by models (i.e., $s_{\pm}$ cases) that imply that the lens system consists of binary stars (see Table \ref{table:model_1243}). However, we find that there exist competing 2L1S models ($\Delta\chi^{2} < 4.8$) that indicate that the lens is likely to be a planetary system (i.e., $q < 0.03$). Indeed, we predict $s^{\dagger}_{-} = 0.990$ and $s^{\dagger}_{+} = 1.011$ from the heuristic analysis ($\tau_{\rm anom} = 0.0067$, $u_{\rm anom} = 0.0211$), which is similar to the $s^{\dagger} \equiv \sqrt{s_{-}s_{+}} = 1.024$ for the combination of P3 and P4 cases. 
In addition, we find that the 1L1S model with the finite-souce effect (FS) has $\Delta\chi^{2} = 9.6$ compared to the best-fit model. However, this 1L1S+FS model is unlikely because the model parameters imply that $\mu_{\rm rel} \sim 0.08\, {\rm mas\, yr^{-1}}$ by assuming a dwarf source (i.e., $\theta_{\ast} \sim 0.5 \mu{\rm as}$) that yields $\mu_{\rm rel} = \theta_{\ast} /(\rho_{\ast}t_{\rm E}) \sim 0.08\, {\rm mas\, yr^{-1}}$). The exceptional small $\mu_{\rm rel}$ is unreliable considering the typical value of bulge/bulge lensing event (i.e., $\mu_{\rm rel} = 5 \sim 10\, {\rm mas\, yr^{-1}}$). 
In Table \ref{table:model_1243}, we present the model parameters of these various competing models. 

\begin{figure}[htb!]
\epsscale{1.00}
\plotone{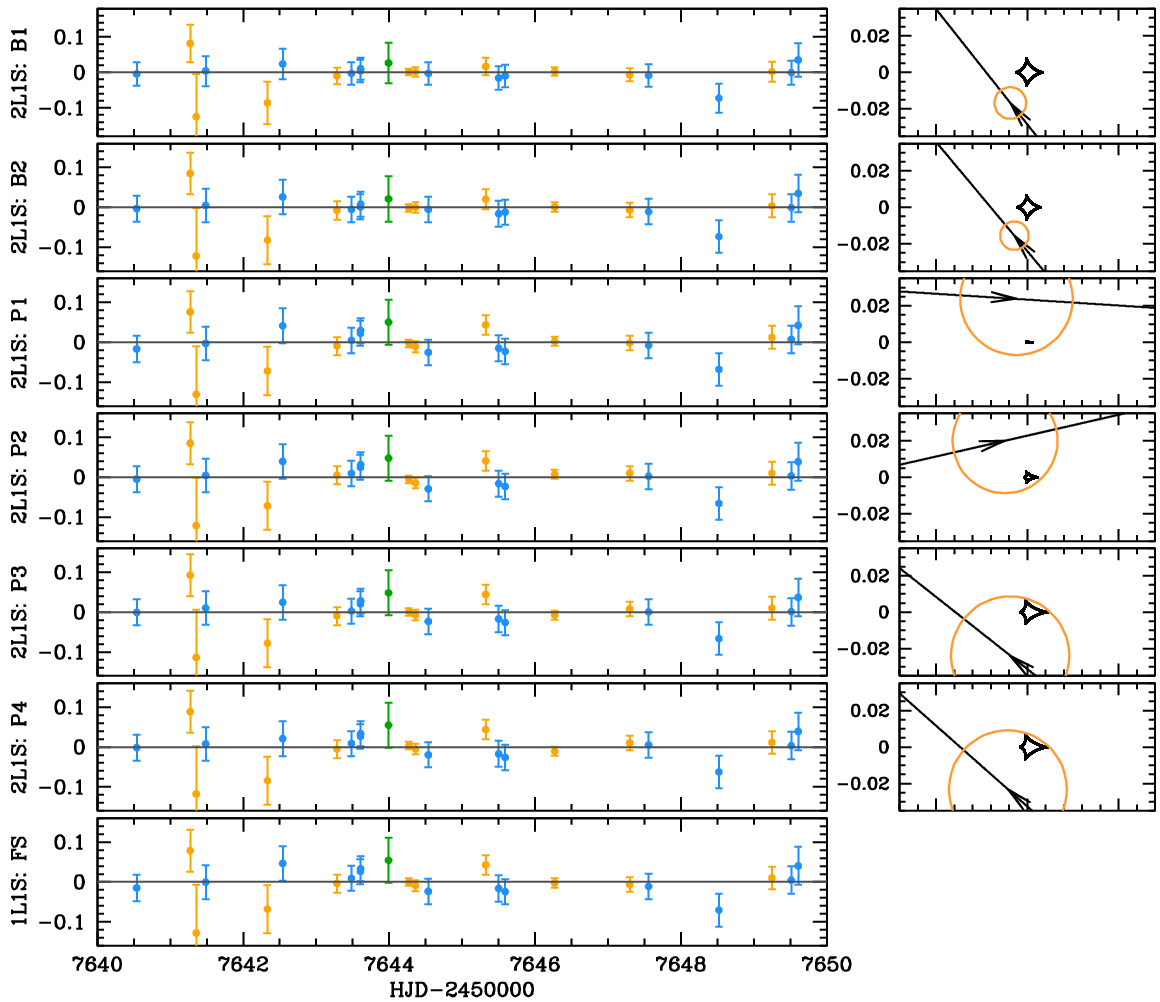}
\caption{\twelvefiftyeight: Residuals of each case shown in Table \ref{table:model_1243} with 
its caustic geometry. We show the residuals for the zoom-in part of Figure \ref{fig:lc_1243_01}. 
\label{fig:lc_1243_02}}
\end{figure}

In Figure \ref{fig:lc_1243_02}, we present the residuals of the anomaly part for all degenerate models with their caustic geometries. By comparing them, we find the $\chi^{2}$ difference mostly comes from fits between HJD$^{\prime} = 7643.5 \sim 7646.0$. However, because of the sparse coverage, the $\Delta\chi^{2}$ of all degenerate cases are smaller than our $\chi^{2}$ criterion ($\Delta\chi^{2} = 10.0$) to claim a planet detection. In particular, the best-fit model of the planet case shows only $\Delta\chi^{2} = 1.8$. Although the binary-lens cases can perfectly describe the peak anomaly (i.e., subtle deviations), there is no clear evidence to rule out the degenerate cases. Thus, we report this event as a planet candidate.

Lastly, we note that we test the APRX effect because of the long timescales (i.e., $t_{\rm E} > 70$ days). However, we find negligible $\chi^{2}$ improvement of $2.8$ compared to the STD best-fit case.

\subsubsection{\threethirtysix}  % Planet Candidate : KMT-2016-BLG-1406 | OGLE-2016-BLG-0336(*) | MOA-2016-BLG-092
\begin{figure}[htb!]
\epsscale{1.00}
\plotone{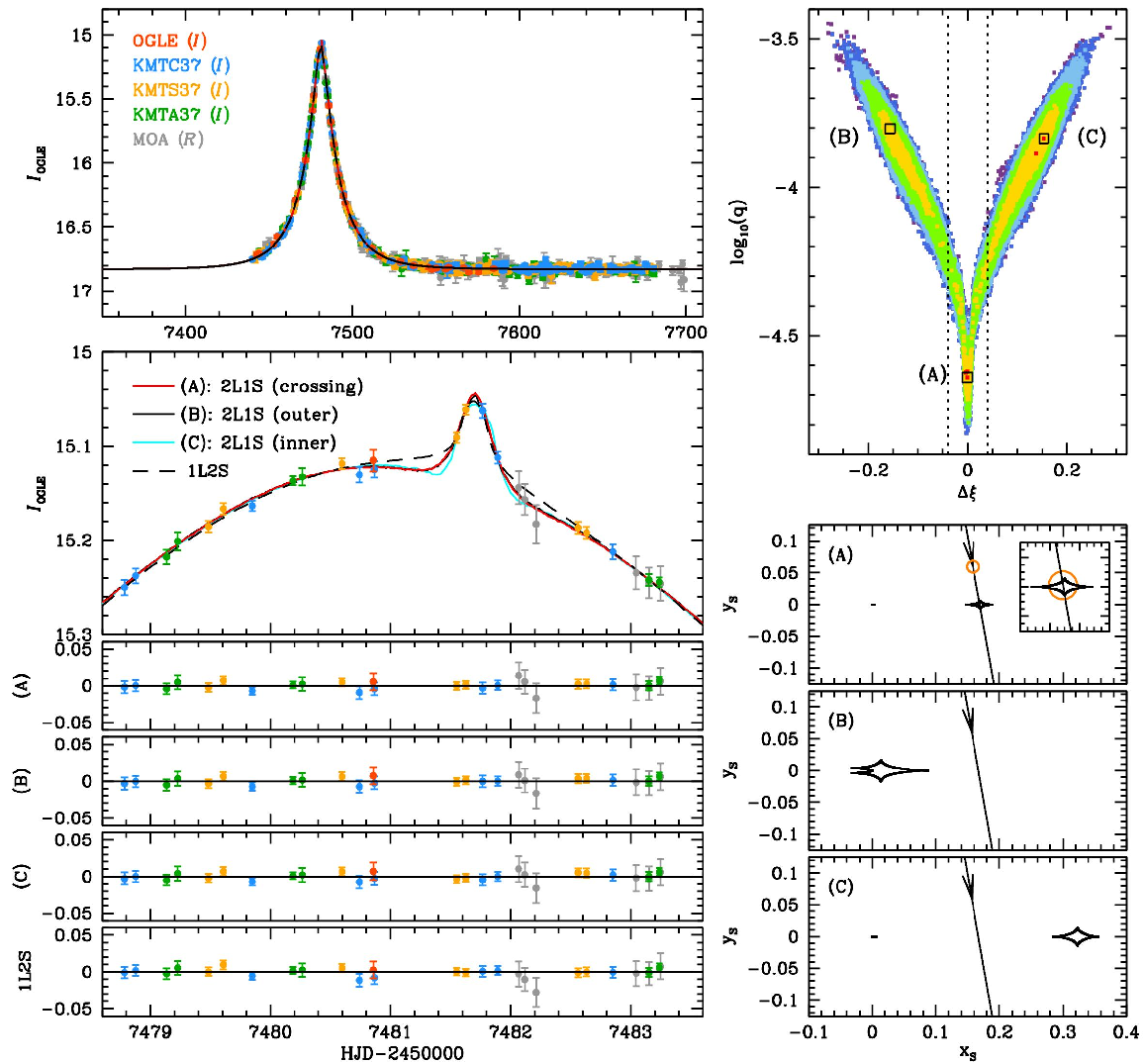}
\caption{Light curves of \threethirtysix\ with degenerate models and their residuals. 
We present $\Delta\xi - \log_{10}(q)$ space to show the local minima for the 2L1S models. 
We also present their caustic geometries.  
\label{fig:lc_1406}}
\end{figure}

As shown in Figure \ref{fig:lc_1406}, the light curve of \threethirtysix\ (which we identified as KMT-2016-BLG-1406) shows an apparent bump-shaped anomaly at the peak (HJD$^{\prime} = 7481.7$), which was covered by KMTC and KMTS observations. We find that the anomaly can be explained by several models presented in Table \ref{table:model_1406}. Similar to the case of \fivetwentysix, there exist three 2L1S solutions caused by different caustic geometries (i.e., (A) caustic-crossing, (B) inner, and (C) outer trajectories). These cases cannot be resolved (i.e., $\Delta\chi^{2} \lesssim 1$). Indeed, we predict $s^{\dagger}_{-} =0.919$ and $s^{\dagger}_{+} = 1.088$ from the heuristic analysis ($\tau_{\rm anom} = 0.0280$, $u_{\rm anom} = 0.1684$). The $s^{\dagger}_{+}$ is well consistent with the best-fit of $s_{+} = 1.089$. We present the $\Delta\xi - \log_{10}(q)$ space to show the locations of these degenerate cases (see the right-upper panel in Figure \ref{fig:lc_1406}). Although we cannot resolve the degeneracy, the mass ratios of all 2L1S solutions imply that the lens is likely to be a planetary lens system (i.e., $q < 0.03$).

However, the bump-shaped anomaly is a typical type to have the 2L1S/1L2S degeneracy. We find that the 1L2S model can describe the anomaly well. Moreover, the $\Delta\chi^{2}$ compared to the 2L1S best-fit model is only $1.13$. Because there are only weak constraints on $\rho_{\ast,\rm S1}$ and $\rho_{\ast,\rm S2}$, and a relatively large separation between the two sources ($\Delta u \sim 0.17$), we cannot place any additional meaningful constraints from physical considerations. Based on currently available data sets and analysis results, we cannot resolve the 2L1S/1L2S degeneracy for this event. Thus, we treat this event as a planet candidate unless we have additional evidence to rule out the 1L2S solution.

Note that we have checked the APRX effect for this event because of the relatively long timescale ($t_{\rm E} \sim 25$ days). We find the $\chi^{2}$ improvement of $14.83$ for the APRX-included model. However, we find that the $\chi^{2}$ improvements between data sets are inconsistent. Indeed, the STD model shows better fits for the KMTC data that yields $\Delta\chi^{2} \sim 10.0$. In contrast, for the other data (OGLE, MOA, and KMTA), the APRX model shows better fits that yield $\Delta\chi^{2} \sim 8.0$, $12.0$, and $4.0$, respectively. For KMTS, there is no $\chi^{2}$ improvement. This inconsistency makes us suspect the APRX detection is unreliable, similar to the case of \fifteenninetyeight. Also, these improvements only come from the baseline, which can have systematics. Thus, we conclude that the APRX measurement is not robust. The STD models should be the fiducial solutions for this event.

\subsubsection{\eighteightytwo} % Planet Candidate : KMT-2016-BLG-1449 | OGLE-2016-BLG-0882(*)
\begin{figure}[htb!]
\epsscale{1.00}
\plotone{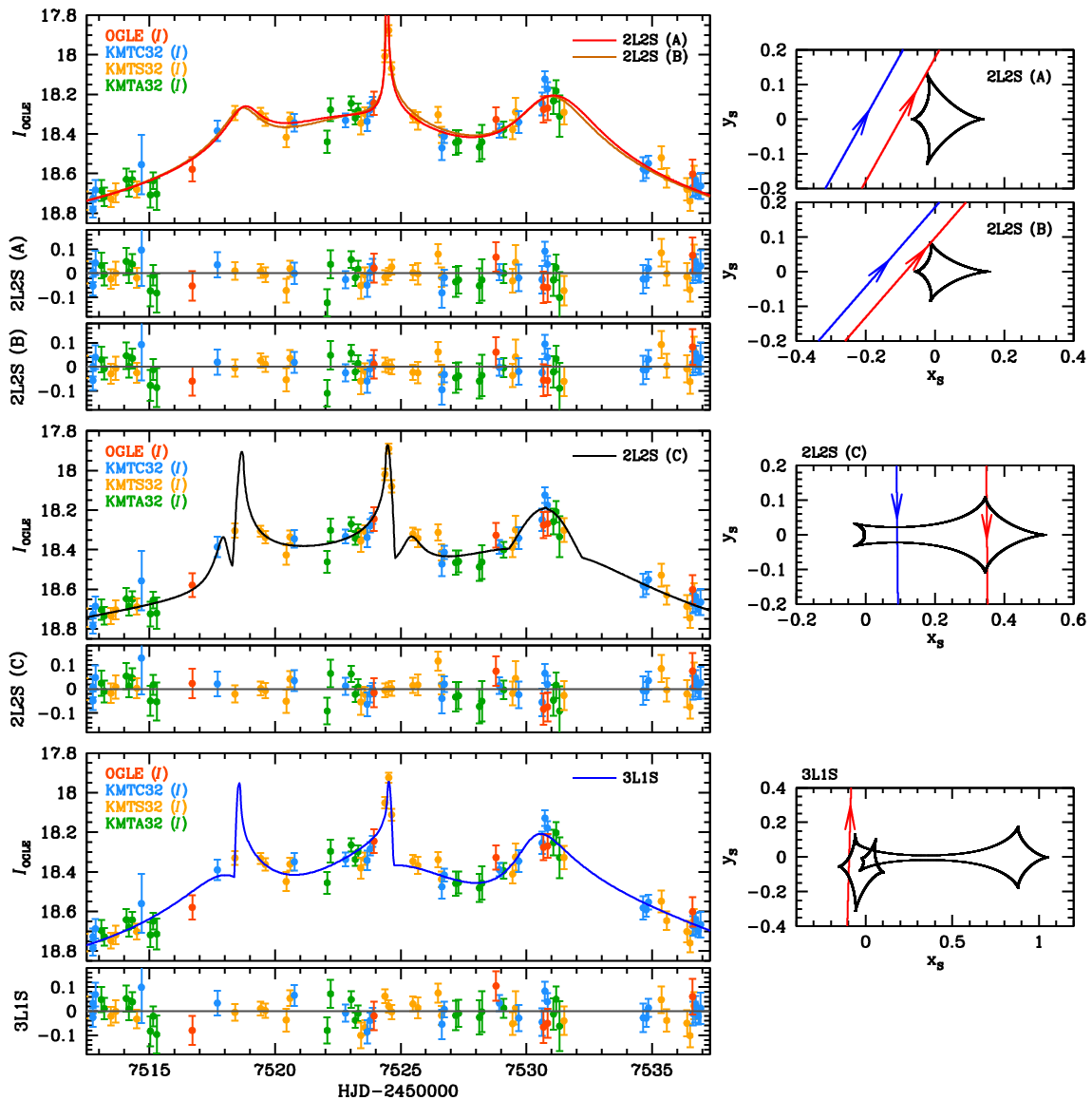}
\caption{Light curves of \eighteightytwo\ with degenerate models and their residuals. 
We also present the caustic geometries of each case.
\label{fig:lc_1449}}
\end{figure}

The light curve of \eighteightytwo\ (which we identified as KMT-2016-BLG-1449) shows anomalies at the peak, which have complex features consisting of three bump-shaped anomalies as shown in Figure \ref{fig:lc_1449}. We find no 2L1S models that can correctly describe the anomalies. Thus, we try to describe the anomalies using 2L2S and 3L1S interpretations. We find the best-fit 2L2S model can describe all anomalies, which implies that the lens system consists of binary stars (i.e., $q \sim 0.3$). However, we also find that there exist competing solutions having $\Delta\chi^{2} < 10.0$. In Table \ref{table:model_1449}, we present these degenerate 2L2S solutions. Among them, one case satisfies our mass ratio criterion for planet detection (i.e., $q \sim 0.01 < 0.03$). However, we find that the best-fit models of all 2L2S cases show inconsistency between the $q_{\rm flux}$ and the ratio of $\rho_{\ast}$ values. Although the $\rho_{\ast}$ measurements are uncertain, the inconsistency implies that the best-fit models may be unreliable. Thus, we investigate chains of each case to find models that satisfy a relation, $q_{\rm flux} \sim (\rho_{\ast, {\rm S1}}/\rho_{\ast, {\rm S2}})^{2}$. We find satisfied 2L2S models that have $(\chi^{2}, q)$ are $(1448.243, 0.273)$, $(1446.316, 0.425)$, and $(1446.476, 0.010)$ for (A), (B), and (C) cases, respectively. These solutions are still unable to resolve (i.e., $\Delta\chi^{2} < 2.0$). Moreover, there still exists the binary/planet degeneracy.

In addition, because the complex anomaly could be described by the 3L1S interpretation, we try to find a possible planetary solution. We find a plausible 3L1S model that can describe the anomalies (see Figure \ref{fig:lc_1449} and Table \ref{table:model_1449}). This 3L1S model implies that the third body is likely to be a planet (i.e., $q_{2} \sim 0.011$). However, this 3L1S model shows $\Delta\chi^{2} > 32.0$ compared to the best-fit 2L2S models. If we consider the satisfied 2L2S models, the 3L1S has worse fits by $\Delta\chi^{2} > 23.0$. Thus, the 3L1S case can be nominally ruled out considering our $\chi^{2}$ criterion. However, we do not ignore the possibility of the 3L1S solution because of two reasons. First, our search for 3L1S models was not exhaustive because of the technical difficulty of a full search of six-dimension parameter space. Thus, there may exist alternative 3L1S solution(s) having better $\chi^{2}$. Second, our $\chi^{2}$ criteria is developed for degeneracies in 2L1S and 1L2S interpretations. Thus, we cannot guarantee the $\chi^{2}$ threshold can be valid for the 2L2S/3L1S degeneracy. Especially for this event, the data sets have systematics on the anomaly part. Hence, we present the 3L1S planetary solution as one alternative possibility of planetary systems that could produce the anomaly. Indeed, if we rule out this 3L1S case, there still remains the binary/planet degeneracy in the 2L2S solutions. Thus, we treat this event as a planet candidate including the possible 3L1S solution.          

Lastly, we note that we test the APRX effect because the models show that the timescales are longer than $32$ days. However, we find only negligible $\chi^{2}$ improvement (i.e., $\Delta\chi^{2} \sim 4.7$) when the APRX effect is considered. Thus, we conclude the STD models are fiducial solutions for this event.

\subsubsection{\seventeenfour} % Planet Candidate : KMT-2016-BLG-1609 | OGLE-2016-BLG-1704(*)
\begin{figure}[htb!]
\epsscale{1.00}
\plotone{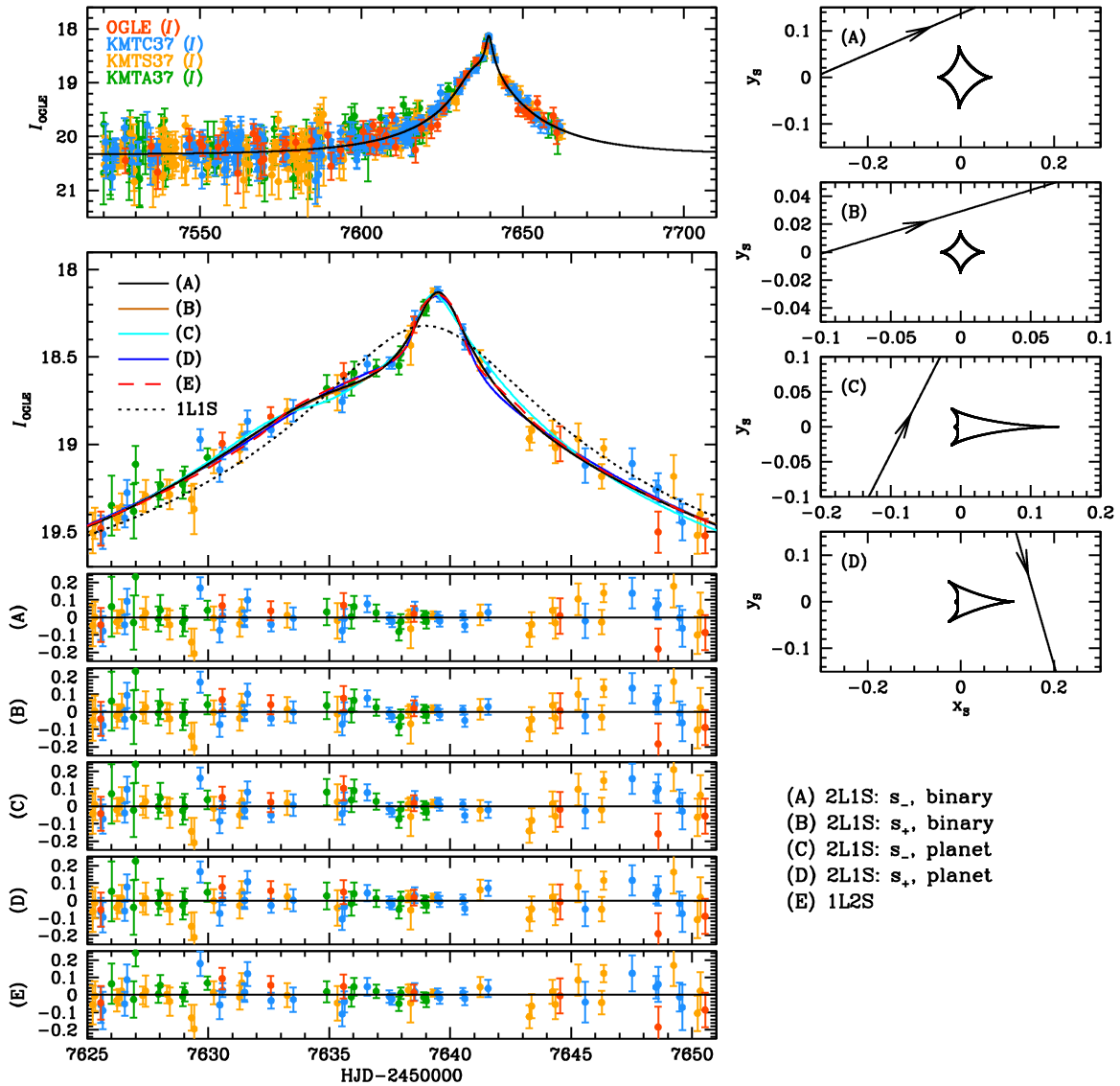}
\caption{Light curves of \seventeenfour\ with degenerate models and their residuals. 
We also present the caustic geometries of each 2L1S case.
\label{fig:lc_1609}}
\end{figure}

The light curve of \seventeenfour\ (which we identified as KMT-2016-BLG-1609) shows apparent deviations from the 1L1S fit. The anomaly can be explained by various models. In Figure \ref{fig:lc_1609}, we present these models with their caustic geometries. As shown in Table \ref{table:model_1609}, the best-fit model (see the (A) case) implies that the lens system consists of binary stars (i.e., $q \sim 0.53$). However, there exist degenerate models having $\Delta\chi^{2} < 10.0$. The mass ratio of the (B) case nominally indicates that the lens is likely to be a binary star system. However, this model is caused by the Chang \& Refsdal lensing \citep{chang79}, which has large uncertainties in the $(s,q)$ parameters. Hence, the mass ratio satisfies our mass ratio criterion (i.e., $q < 0.03$) within $1\sigma$. For the (C) case, the mass ratio indicates the lens system could have a planet. The (D) solution can be nominally resolved by $\Delta\chi^{2} = 13.1$, which is slightly larger than our $\chi^{2}$ criterion. However, by considering the systematics in the data sets, we cannot firmly rule out this case. Thus, we present this planet-like case for completeness. For the (C) and (D) cases, the heuristic analysis ($\tau_{\rm anom} = 0.0217$, $u_{\rm anom} = 0.0752$) predicts $s^{\dagger}_{-} = 0.963$ and $s^{\dagger}_{+} = 1.038$, which is similar to the empirical value of $s^{\dagger} = \sqrt{s_{-,{\rm (C)}}s_{+,{\rm (D)}}} = 1.035 $. 

Lastly, we find that a 1L2S model can also explain the anomaly. The $\Delta\chi^{2}$ between the best-fit and 1L2S models is only $3.4$, which cannot be resolved. Thus, we treat this event as a planet candidate because of the binary/planet and 2L1S/1L2S degeneracies.

We note that we have tested the APRX effect because of the relatively long timescales (i.e., $t_{\rm E} > 32$ days). We find the negligible $\chi^{2}$ improvement of $5.0$ when the APRX effect is included. Thus, we conclude that the STD models are the fiducial solutions for this event.

\subsubsection{\fourteeneight} % Planet Candidate : KMT-2016-BLG-1630 | OGLE-2016-BLG-1408(*)
\begin{figure}[htb!]
\epsscale{1.00}
\plotone{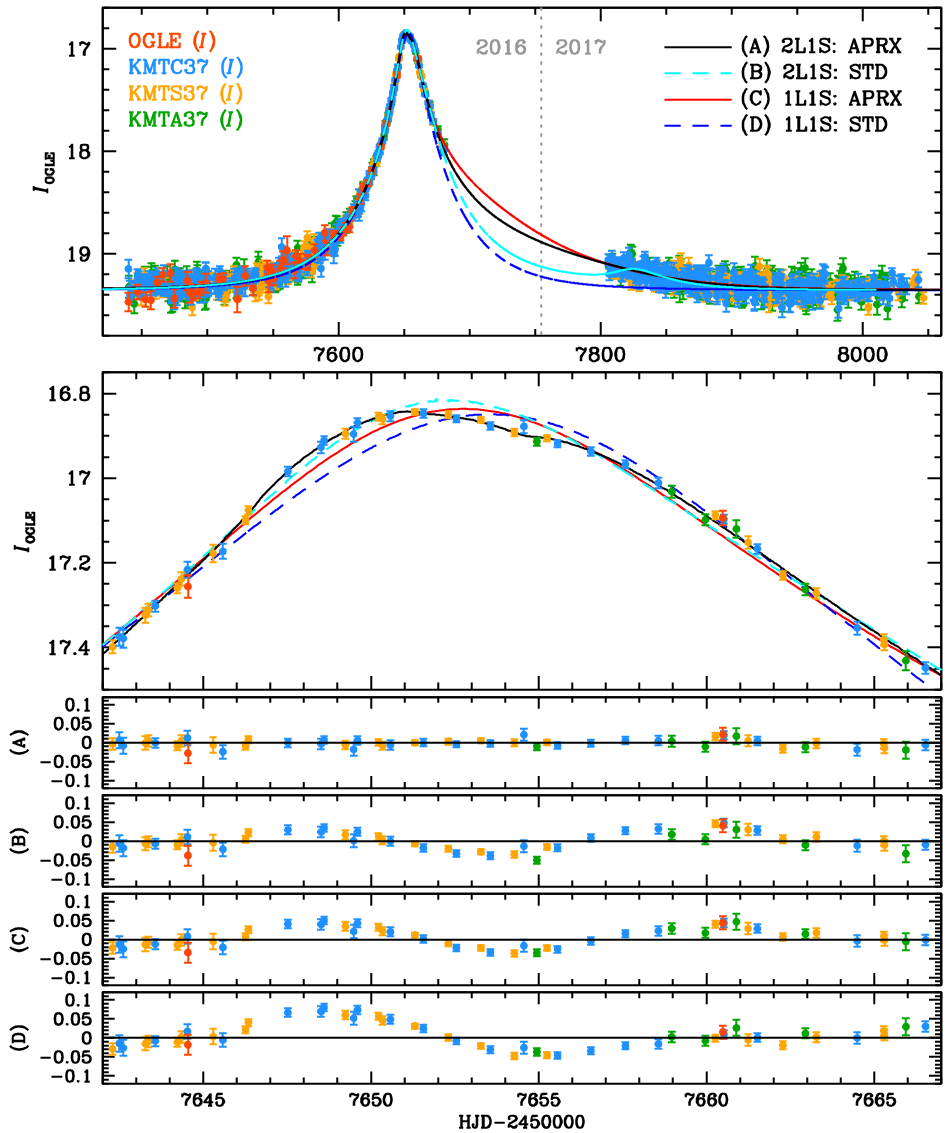}
\caption{Light curves of \fourteeneight\ with 2L1S and 1L1S models. 
For the 2L1S and 1L1S models, we present both the STD and APRX cases.  
\label{fig:lc_1630_01}}
\end{figure}

\fourteeneight\ (which we identified as KMT-2016-BLG-1630) is a long timescale event that has an anomaly at the peak on the light curve. In Figure \ref{fig:lc_1630_01}, we present the light curve with the 2L1S and 1L1S models of the STD and APRX cases. Because of the long timescale (i.e., $t_{\rm E} > 96$ days), we find that the APRX effect is essential to describe the observed light curve. In particular, as shown in Figure \ref{fig:lc_1630_01}, it is impossible to describe the 2017 data without the APRX effect. Also, the 2L1S models with the APRX effect are the only interpretations that can explain the anomaly at the peak.

\begin{figure}[htb!]
\epsscale{1.00}
\plotone{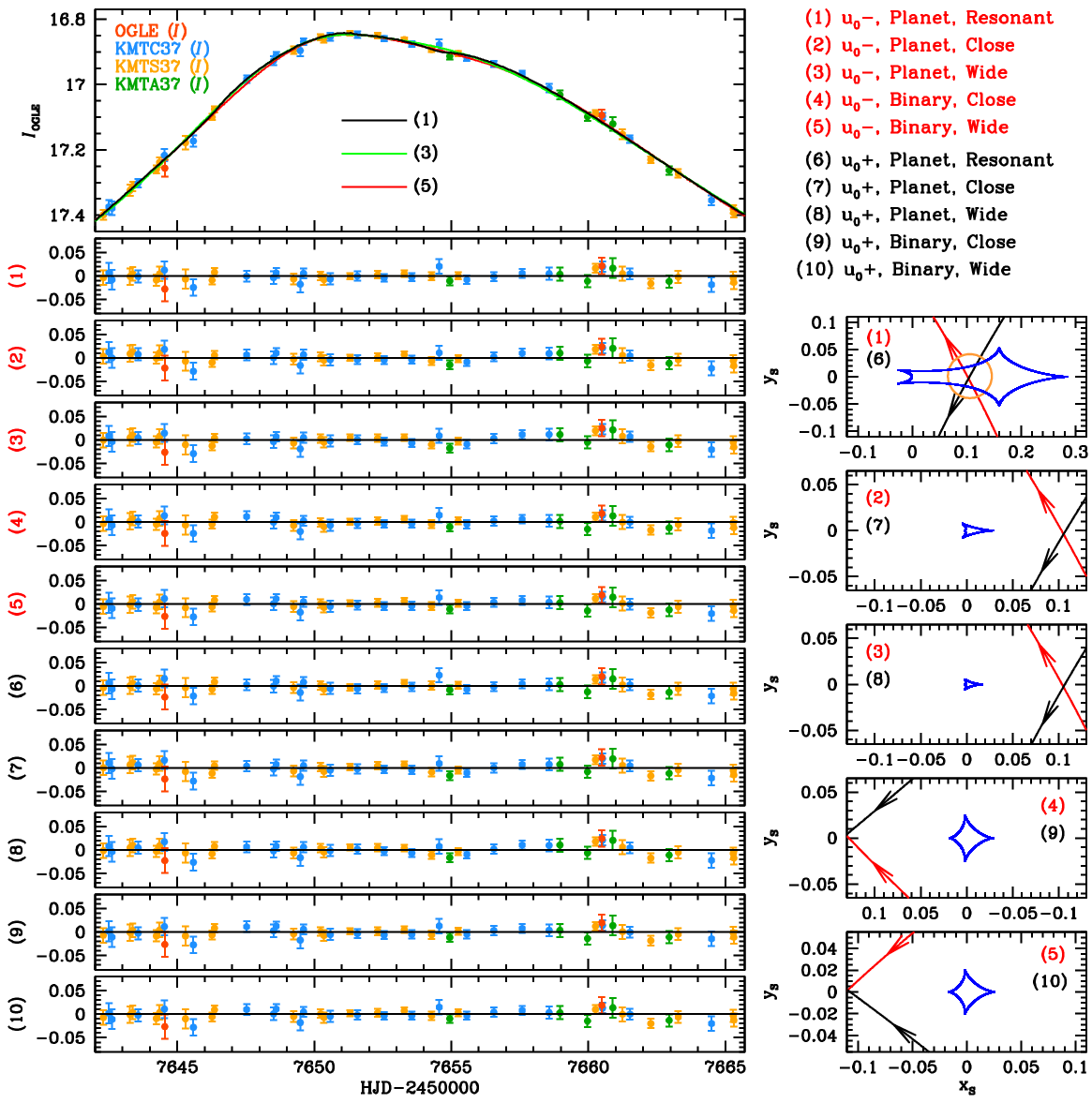}
\caption{Comparison of 2L1S APRX models with their caustic geometries for \fourteeneight. 
\label{fig:lc_1630_02}}
\end{figure}

However, we find that several 2L1S APRX models can describe the whole light curve, which cannot be distinguished from each other. In Figure \ref{fig:lc_1630_02}, we present these degenerate solutions with their caustic geometries. We also present model parameters for the cases in Table \ref{table:model_1630}. The best-fit case indicates that the lens could be a planetary system (i.e., $q \sim \mathcal{O}(10^{-3})$). There exist five competing planetary cases caused by the close/wide \citep{griest98} and ecliptic \citep{smith03, jiang04,poindexter05} degeneracies. Although among the planetary cases, the wide $u_{0}\pm$ cases can be nominally resolved by $\Delta\chi^{2} > 10.0$, we present them for completeness and comparison to the binary-lens cases. 

Despite the best-fit model implying that the lens has a planet, we find that there also exist competing binary-lens cases having $\Delta\chi^{2} \lesssim 5.4$. In particular, the best fit of the binary case shows only $\Delta\chi^{2} = 0.9$. Thus, we treat this event as a planet candidate because we cannot resolve the planet/binary degeneracy.

\begin{figure}[htb!]
\epsscale{0.90}
\plotone{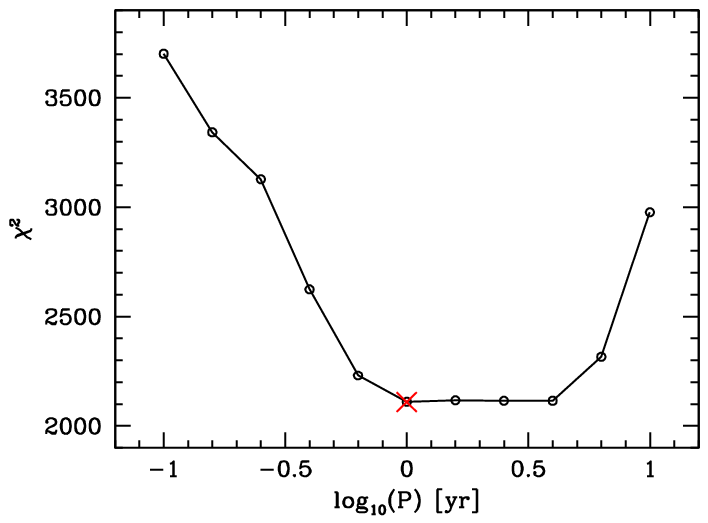}
\caption{Test of the xallarap effect for \fourteeneight, which shows $\chi^{2}$ value for each rotation period ($P$) of 
the binary source system. The red cross indicates the best-fit $\chi^{2}$ of the APRX model, which is equal to one year. 
\label{fig:lc_1630_xallarap}}
\end{figure}

We note that we conduct tests for the APRX effect because the effect is essential to finding the solutions. First, we have tested the OBT effect, which can affect the APRX measurement. We find no $\chi^{2}$ improvement when the OBT effect is considered (i.e., $\Delta\chi^{2}\, [\rm {OBT-APRX}]= 0.3$). Moreover, we find that the OBT effect does not affect the uncertainties of the APRX measurement. Second, we have tested whether the xallarap effect can mimic the APRX effect. Similar to the OBT case, we find that the xallarap effect does not improve the fits (i.e., $\Delta\chi^{2}\, [\rm {xallarap - APRX}]= 0.4$). Also, as shown in Figure \ref{fig:lc_1630_xallarap}, the best-fit xallarap model has $P = 1$ yr, which is consistent with the orbital period of the Earth. Both facts imply that the effect on the light curve is caused by APRX rather than xallarap. Hence, we conclude that the APRX models are the fiducial solutions for this event. We also note that we can measure $\rho_{\ast}$ for only the resonant ($u_{0}\pm$) cases induced by the caustic-crossing geometries. For other cases, we cannot robustly measure the $\rho_{\ast}$ because of the non-caustic-crossing geometries.

Lastly, we check the 2L1S/1L2S degeneracy because the bump-like anomaly can be explained by the 1L2S interpretation. We find that the 1L2S model with the APRX effect shows better fits by $\Delta\chi^{2} = 6.69$ compared to the best-fit of the 2L1S APRX models. However, the 1L2S model is not reliable because the $q_{\rm flux}$ and a ratio of $\rho_{\ast}$ values are inconsistent. In addition, there is no case to satisfy the relation, $q_{\rm flux} \propto (\rho_{\ast, {\rm S1}}/\rho_{\ast, {\rm S2}})^{2}$, in the all chains. Thus, we can rule out the 1L2S model although it shows better fits.

\subsubsection{\twentythreeninetynine}  % Planet Candidate : KMT-2016-BLG-2399
\begin{figure}[htb!]
\epsscale{1.00}
\plotone{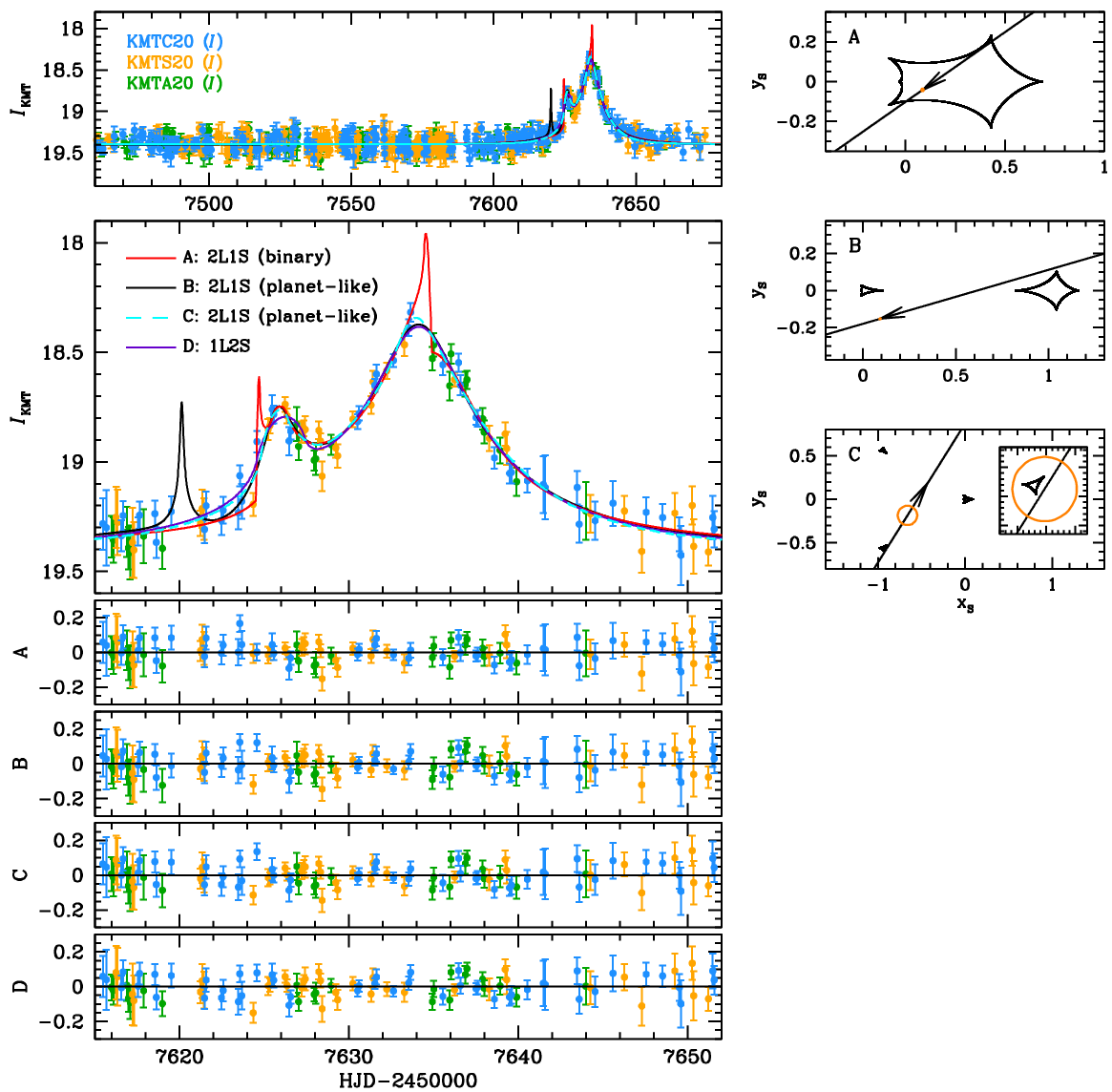}
\caption{Light curves of \twentythreeninetynine\ with 2L1S and 1L2S models. Note model B has a sharp feature near 
HJD$^{\prime}\sim7620$ that lacks data coverage. We also present the caustic geometries of the 2L1S models.
\label{fig:lc_2399}}
\end{figure}

The light curve of \twentythreeninetynine\ shows a bump-shaped anomaly on the rising part (HJD$^{\prime} \sim 7626$). As shown in Figure \ref{fig:lc_2399}, the anomaly can be described by a binary-lens model that contains a low-mass object (i.e., $q \sim 0.057$). We also find that planet-like models can plausibly describe the anomaly. In Table \ref{table:model_2399}, we present the model parameters of possible solutions for this event. Indeed, the heuristic analysis ($\tau_{\rm anom} = -0.2813$, $u_{\rm anom} = 0.2924$) predicts $s^{\dagger}_{-} = 0.864$ and $s^{\dagger}_{+} = 1.157$, which is consistent with $s^{\dagger} \equiv \sqrt{s_{+}(A)s_{-}(C)} = 0.864$. In addition, we find that the bump-shaped anomaly can also be plausibly described by a 1L2S model, which shows $\Delta\chi^{2} = 14.2$ compared to the best-fit model.

We note that the planet-like cases are borderline given our criteria. 
First, for the B case, the mass ratio is $\sim 0.030$, which is consistent with the $q$ criterion, while the C case does not satisfy the $q$ criterion. However, the C model shows a very short timescale (i.e., $t_{\rm E} \sim 8$ days) with a relatively small $q$ value (i.e., $q \sim 0.049$), which implies the component of the lens system would be a planet. Second, both cases are nominally resolved by the $\chi^{2}$ criterion (i.e., $\Delta\chi^{2} = 10.0$). However, the B case ($\Delta\chi^{2} = 10.2$) is very close to our $\chi^{2}$ criterion. By considering the systematics in the data, we cannot firmly rule out the model based on current data. We note that the B model exhibits a sharp bump at HJD$^{\prime} \sim 7620$. However, there are no available data points observed by either KMTNet or OGLE. 

Even if we can rule out the C and 1L2S cases by simply adopting our criteria, there still remains a possible planet case (i.e., the B case) that cannot be clearly ruled out. Thus, we treat this event as a planet candidate. 

Note that we have tested the APRX effect for this event because the best-fit solution has a sufficiently long timescale (i.e., $t_{\rm E} \sim 19$ days) that the APRX effect may be detected. However, we find a negligible $\chi^{2}$ improvement of $0.9$ when we consider the APRX effect. Thus, the STD cases are the fiducial models for this event. Finally, we note that we can measure the $\rho_{\ast}$ values for the A (caustic-crossing) and C (buried caustic) cases (see caustic geometries in Figure \ref{fig:lc_2399}).

\subsubsection{\twentyfourseventythree} % Planet Candidate : KMT-2016-BLG-2473  

\begin{figure}[htb!]
\epsscale{1.00}
\plotone{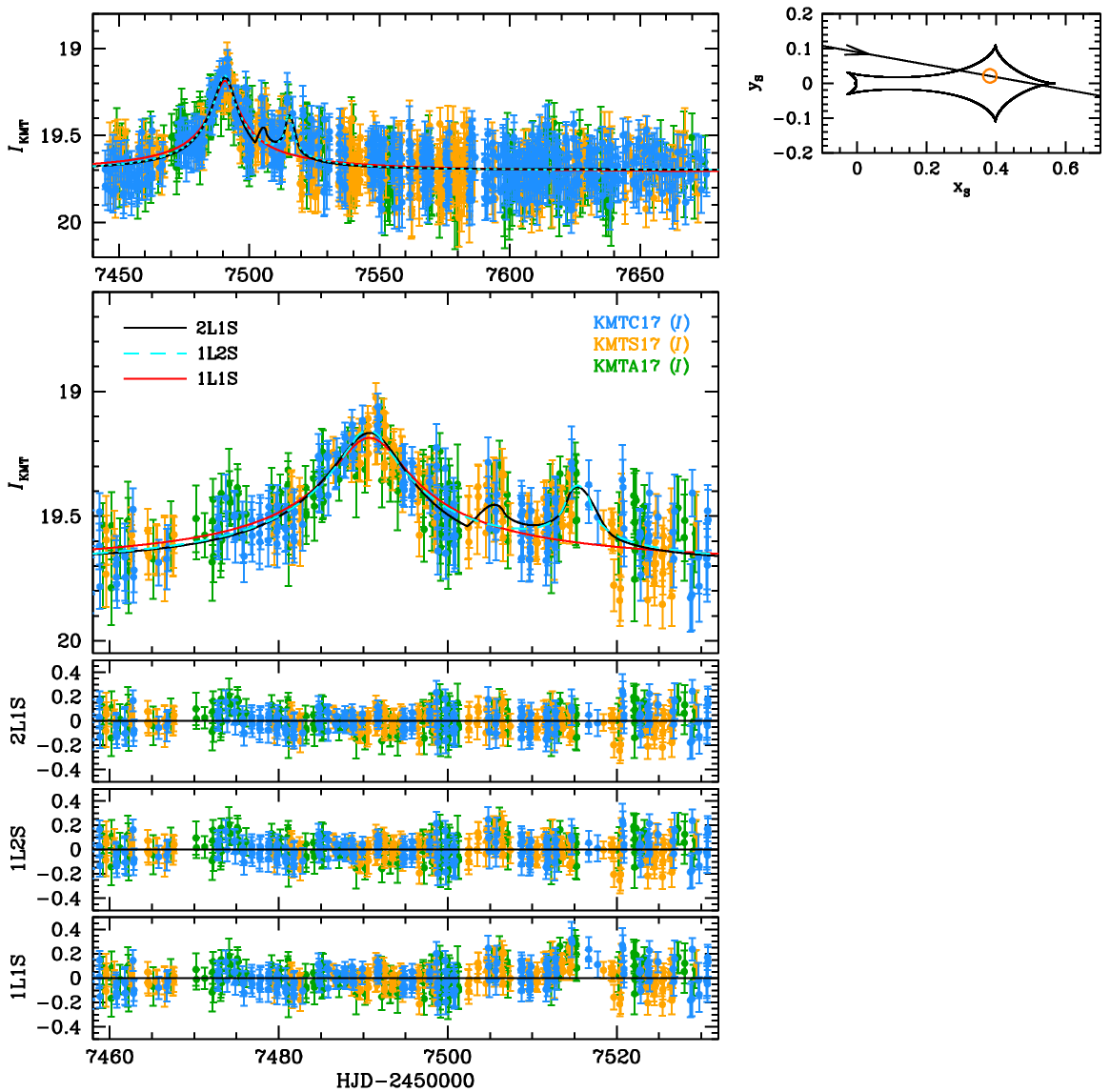}
\caption{Light curves of \twentyfourseventythree\ with 2L1S and 1L2S models, their residuals, and the 2L1S caustic geometry.
\label{fig:lc_2473}}
\end{figure}

The light curve of \twentyfourseventythree\ exhibits anomalies from the 1L1S model ($\Delta\chi^{2} = 171.0$) during HJD$^{\prime} = 7500 \sim 7520$, as shown in Figure \ref{fig:lc_2473}. The anomalies can be explained by a 2L1S model (note that the heuristic analysis is not valid for this event). The mass ratio of this best-fit model indicates that the lens system is likely to be a planetary system (i.e., $q \sim 0.011$). However, we find that a 1L2S model is also able to plausibly describe the anomaly. In Table \ref{table:model_2473}, we present the model parameters of the 2L1S and 1L2S models.

The 2L1S and 1L2S models themselves show a clear difference at HJD$^{\prime} \sim 7505.0$, which seems to be a shallow bump-shaped anomaly. However, the $\Delta\chi^{2}$ between them is only $10.3$, which does not satisfy our criterion to resolve the 2L1S/1L2S degeneracy. The small $\Delta\chi^{2}$ is caused by severe systematics in data sets because the event experienced heavy extinction (i.e., $A_{I} \sim 4.9$). Thus, we treat this event as a planet candidate because we do not have any conclusive evidence to resolve the 2L1S/1L2S degeneracy.

Note that we have tested the APRX effect because of the long timescale (i.e., $t_{\rm E} \sim 47$ days). We find a small $\chi^{2}$ improvement of $5.8$ when we consider the APRX effect. However, the improvement comes from the baseline, which has severe systematics. Thus, we conclude that the APRX effect is not robust. Hence, the STD models are the fiducial solutions for this event. Lastly, we note that the $\rho_{\ast}$ can be measured for the 2L1S case from the caustic-crossing feature.

\section{CMD Analysis} \label{sec:CMDs}

We cannot securely measure $\rho_{\ast}$ for any of the four planetary events. We can determine only upper limits on the $\rho_{\ast}$ values. However, we can apply the $\rho_{\ast}$ distributions as constraints on the Bayesian analysis by including information on the angular source radius ($\theta_{\ast}$) of each event in the analysis. Thus, we carry out the color-magnitude diagram (CMD) analysis to measure the $\theta_{\ast}$. The basics of the CMD analysis are described in \citet{yoo04}. In addition, the detailed procedures of the analysis are described in \citet{shin23b}. 

\begin{figure}[htb!]
\epsscale{1.00}
\plotone{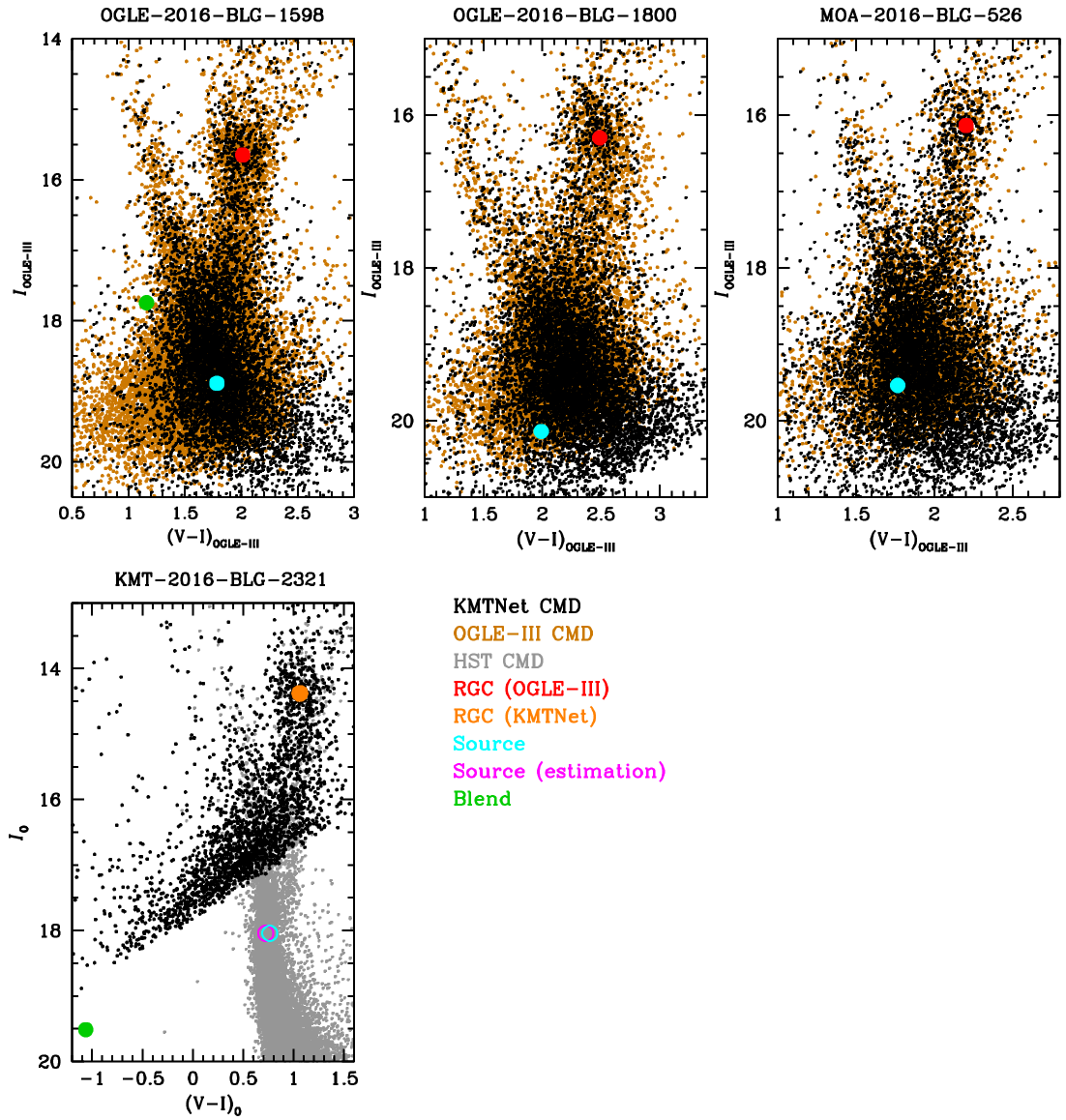}
\caption{Color--magnitude diagrams (CMDs) of four planetary events.
\label{fig:CMDs}}
\end{figure}

In Figure \ref{fig:CMDs}, we present the measured locations of the centroid of the red giant clump (RGC), the source, and the blend overlaid on the CMD of each event. Although the analysis is conducted based on the multi-band KMTNet observations (i.e., $I$ and $V$ bands), we present them in the OGLE-III magnitude system because we determine the RGC based on the OGLE-III CMD \citep{szymanski11}. The exception is \twentythreetwentyone\ because the OGLE-III CMD is not available for this event, so we present the uncalibrated/dereddend KMTNet magnitudes instead. 

In Table \ref{table:cmd}, we present the results of the CMD analyses with the derived $\theta_{\ast}$ values. We also present the lower limits on the angular Einstein ring radii ($\theta_{\rm E}$) and lens-source relative proper motions ($\mu_{\rm rel}$). Indeed, the lower limit on $\mu_{\rm rel}$ (i.e., $\mu_{{\rm rel}, +3\sigma} \equiv \theta_{\ast}/t_{\rm E}\rho_{\ast,+3\sigma}$) is a useful indicator to check the effect of the $\rho_{\ast}$ constraint before proceeding with the actual Bayesian analysis. In general, we expect $1 < \mu_{\rm rel} / {\rm mas\, yr^{-1}} < 10$. Hence, if the lower limit on $\mu_{{\rm rel}, +3\sigma} \lesssim 1\, {\rm mas\, yr^{-1}}$, we expect the $\rho_{\ast}$ constraint to have little effect on Bayesian result.

Note that, for \twentythreetwentyone, we conduct additional analysis to check our measurement of the source color because the quality of the $V$-band data is low. The $V$-band light curve has systematics because this event experienced severe extinction (i.e., $A_{I} \sim 3.88$) and the source is faint (i.e., $I_{\rm KMTNet} \sim 21.4$). Thus, we have checked our measurement using the source color estimation method \citep{bennett08} and the Galactic bulge CMD \citep{holtman98} from the Hubble Space Telescope (HST). We find that the estimated source color (i.e., $(V-I)_{0,{\rm S}} = 0.723 \pm 0.055$) is consistent with our measured color (i.e., $(V-I)_{0,{\rm S}} = 0.763 \pm 0.092$ or $(V-I)_{0,{\rm S}} = 0.752\pm0.090$) at the $1\sigma$ level. Hence, we conclude that our measurement is reliable despite the obstacles.

\section{Planet Properties} \label{sec:lens_properties}
The lens properties such as the mass of the lens system ($M_{\rm L}$), distance to the lens ($D_{\rm L}$), projected separation between lens components ($a_{\perp}$), and lens-source relative proper motion ($\mu_{\rm rel}$) can be determined from
\begin{equation}
    M_{\rm L} = \frac{\theta_{\rm E}}{\kappa \lvert \pivec \rvert},~~ 
    D_{\rm L} = \frac{\rm au}{\theta_{\rm E}\lvert \pivec \rvert + \pi_{\rm S}},~~
    a_{\perp} = sD_{\rm L}\theta_{\rm E}, ~~
    \mu_{\rm rel} = \frac{\theta_{\rm E}}{t_{\rm E}},
    \label{eqn:lens}
\end{equation}
where $\kappa = 8.144\, {\rm mas}\,M_{\odot}^{-1}$ and $\pi_{\rm S}$ is the parallax of the source defined as $\pi_{\rm S} \equiv {\rm au}/D_{\rm S}$ ($D_{\rm S}$ is distance to the source). As shown in Equation \ref{eqn:lens}, two observables (i.e., $\theta_{\rm E}$ and $\lvert \pivec \rvert$) need to be measured to directly determine the lens properties. These observables may be measured from the finite-source and microlens-parallax effects, respectively. However, for the planetary events in this work, we do not have measurements of either observable. Thus, we conduct a Bayesian analysis to estimate the lens properties for the new planetary systems. We follow the formalism and procedures of the Bayesian analysis described in \citet{shin23a, shin23b}. 

In Table \ref{table:lens}, we present the lens properties estimated from the Bayesian analyses for each event. Note that we apply the $t_{\rm E}$ and $\rho_{\ast}$ distribution constraints to the Bayesian analyses for all planetary events. For each event, we present several lens properties because of the degenerate solutions. Thus, we present ``adopted" values for ease of cataloging, which are weighted average values described in \citet{jung23}.

\subsection{\fifteenninetyeight} % KMT-2016-BLG-0696 == OGLE-2016-BLG-1598(*) == MOA-2016-BLG-521
The planetary lens system of this event consists of a sub-Jupiter-mass planet ($M_{\rm planet} \sim 0.37$ or $\sim 0.70\, M_{\rm J}$) orbiting an early M-dwarf host star ($M_{\rm host} \sim 0.55, M_{\odot}$) with a projected separation of $\sim 2.5$ or $\sim 1.9$ au. This planetary system is located at a distance of $\sim 5.9$ kpc from us. The properties of the planetary system are those of a typical microlensing planet, i.e., a Jupiter-class planet orbiting an M-dwarf host beyond the snow line \citep{ida05,kennedy08}.

\subsection{\eighteendoubleO} % KMT-2016-BLG-0781 == OGLE-2016-BLG-1800(*) == MOA-2016-BLG-581
For this event, the lens system is composed of a super Jupiter-mass planet ($M_{\rm planet} \sim 2.49$ or $\sim 2.77\, M_{\rm J}$) and an M-dwarf host star ($M_{\rm host} \sim 0.41, M_{\odot}$). The planet orbits the host with a projected separation of $\sim 1.5$ or $\sim 2.6$ au. The system is located at a distance of $\sim 6.5$ kpc from us. This planetary system is also one that is typical for microlensing planets.

\subsection{\fivetwentysix} % KMT-2016-BLG-1611 == OGLE-2016-BLG-1705 == MOA-2016-BLG-526(*)
Despite several solutions, the Bayesian results indicate that the properties of the host star are consistent, i.e., it is an M-dwarf star with the mass of $\sim 0.4\, M_{\odot}$. However, because of the variation in mass ratios for different solutions, the planet could be either a sub-Neptune-mass or Neptune-class planet (i.e., from $M_{\rm planet} \sim 0.3$ to $\sim 1.3\, M_{\rm N}$). This planet orbits its host within a range of the projected separations from $a_{\perp} \sim 2.0$ to $\sim 2.3$ au, where the uncertainty in $a_{\perp}$ is caused by the variation in projected separations for different solutions. This planetary system is located at a distance of $\sim 6.9$ kpc from us.

\subsection{\twentythreetwentyone} % KMT-2016-BLG-2321(*) == no OGLE == no MOA
Bayesian results show that the lens system of this event consists of a Jupiter-class planet ($M_{\rm planet} \sim 0.94$ or $\sim 0.98\, M_{\rm J}$) orbiting a mid-K-type host star ($M_{\rm host} \sim 0.73, M_{\odot}$) with a projected separation of $\sim 3.4$ or $\sim 3.8$ au. The system is located at the distance of $\sim 3.6$ or $\sim 3.5$ kpc. 

Note that, for this event, the constraints from the $\rho_{\ast}$ distributions have a major effect on the posteriors, in contrast to the other cases presented above. Indeed, we can expect the effect of the $\rho_{\ast}$ constraints to be significant as described in Section \ref{sec:CMDs}. Specifically, for this event, $\mu_{\rm rel,{+3\sigma}} \sim 4\, {\rm mas\, yr^{-1}}$, which is much larger than $1\, {\rm mas\, yr^{-1}}$. Meanwhile, for the other events, the effects of the $\rho_{\ast}$ constraints were minor, as would be expected from lower limits of $\mu_{\rm rel,{+3\sigma}} \lesssim 1\, {\rm mas\, yr^{-1}}$ (see Table \ref{table:cmd}).

\section{Summary and Discussion} \label{sec:summary}

Through our systematic planetary anomaly search, we found four hidden planets and seven planet candidates in the 2016 KMTNet sub-prime fields. The properties of these new planetary systems are those of typical microlensing planets, i.e., giant planets orbiting M dwarf host stars beyond their snow lines. Although these new planets show typical properties discovered by the microlensing method, these are complementary planet samples compared to samples discovered by other detection methods because of the different detection sensitivities of each method \citep{clanton14a, clanton14b, shin19}.  

In Table \ref{table:2016_planets}, we present all planetary events observed in 2016, including the new planets of this work. Both the by-eye and the AF methods were used to identify these planets. This work contributes  $31\%$ of the total number of planets discovered in the 2016 KMTNet sub-prime fields. Similarly to the contribution of this work, \citet{shin23b} reported 5 planets, which contributed $33\%$ of the total number of planets discovered in 2016 in the prime fields. Hence, for the high- and low-cadence fields, we found a similar fraction of hidden planets.  

Despite the number of new planets in both fields being similar, the number of new planet candidates shows a big difference. \citet{shin23b} found only one planet candidate in the high-cadence fields. By contrast, we found seven planet candidates in the low-cadence fields. These events are treated as planet candidates because we cannot resolve the binary/planet or 2L1S/1L2S degeneracy, which is caused by non-optimal coverage of the anomalies. This fact clearly shows the importance of high-cadence observations to conclusively claim planet detections. 

\begin{figure}[htb!]
\epsscale{1.00}
\plotone{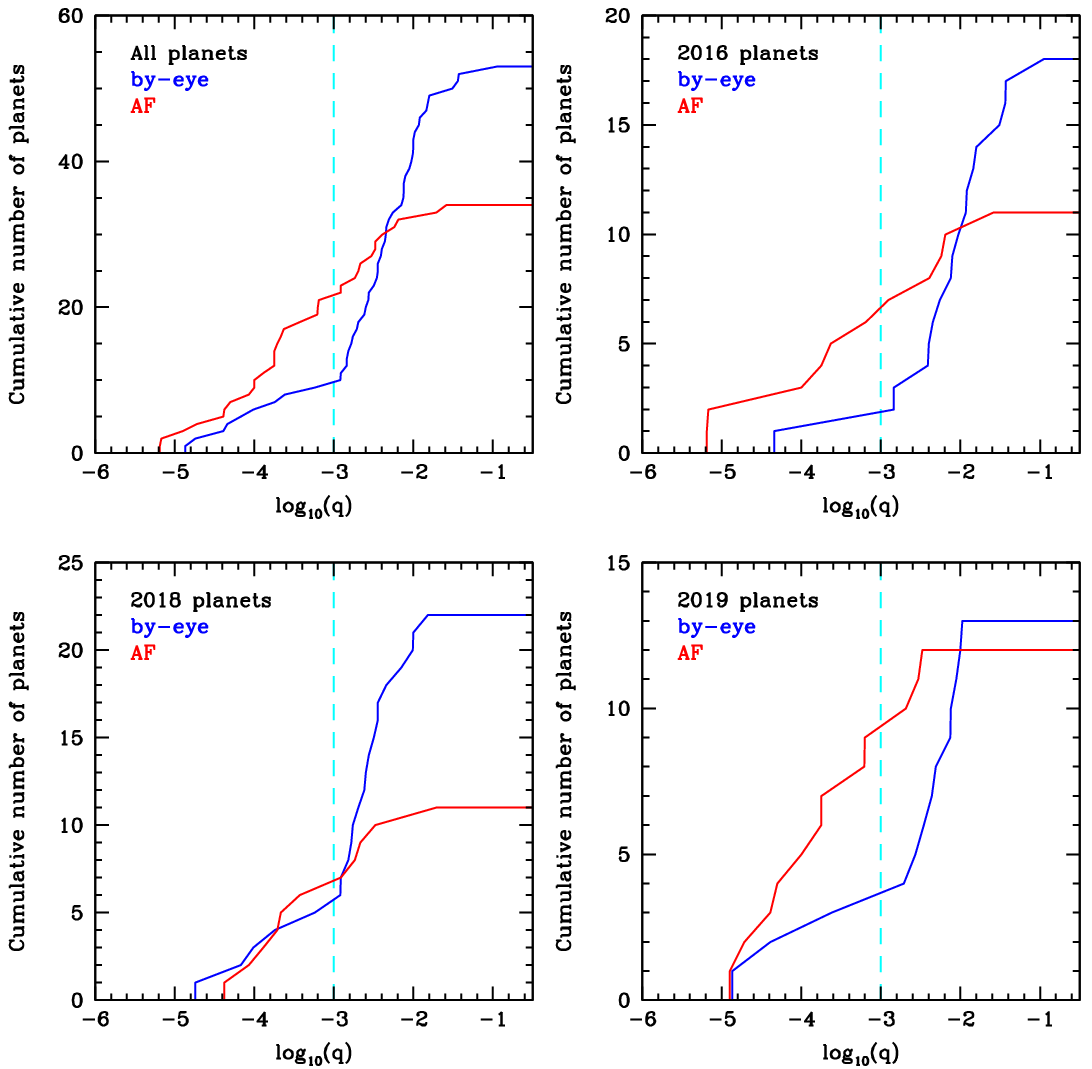}
\caption{Cumulative number of planets discovered by the AF and eye as a function of $\log_{10}(q)$. 
We present $2016$, $2018$, and $2019$ cases that have finished the systematic search for both prime 
and sub-prime fields.
\label{fig:cum_planets}}
\end{figure}

Now that we have finished the systematic search work for both prime and sub-prime fields observed in 2016, 2018, and 2019, in Figure \ref{fig:cum_planets}, we present the cumulative number of planets discovered by the AF and by-eye as functions of $\log_{10}(q)$. For each year, we find that $86\% (=6/7)$, $55\% (=6/11)$, and $75\% (=9/12)$ of total planetary events having $\log_{10}(q) < -3.0$ were identified by the AF method, respectively. Combining the three seasons, a $70\% (=21/30)$ of planetary systems in the region of $\log_{10}(q) < -3.0$ were discovered by the AF method rather than by-eye. This is a remarkable result. Indeed, a total of 53 planetary events were identified by the conventional method (i.e., by eye) in the 2016, 2018, and 2019 seasons. However, only $17\% (=9/53)$ of those planetary systems have $\log_{10}(q) < -3.0$. This lack of planet abundance in the region of $\log_{10}(q) < -3.0$ is unexpected considering the fact that microlensing detections are only weakly dependent on the mass of the planet ($\propto q^{1/2}$) and KMTNet's near-continuous observations should easily capture, e.g., the $\sim 8\, $ hr signals due to $\log_{10}(q) \sim -4$ planets. However, this investigation simply shows that most of the planetary systems having $\log_{10}(q) < -3.0$ were just buried in the archive and missed by by-eye searches. This fact clearly shows the importance of our systematic search to building a complete microlensing planet sample.

\hbox{}
% =================================
% Acknowledgments
% =================================
%\begin{acknowledgments}
% KMTNet
This research has made use of the KMTNet system operated by the Korea Astronomy and Space Science Institute (KASI) at three host sites of CTIO in Chile, SAAO in South Africa, and SSO in Australia. 
Data transfer from the host site to KASI was supported by the Korea Research Environment Open NETwork (KREONET). 
This research was supported by KASI under the R\&D program (project No. 2023-1-832-03) supervised by the Ministry of Science and ICT.
% MOA
The MOA project is supported by JSPS KAKENHI Grant Number JP24253004, JP26247023, JP16H06287 and JP22H00153.
% Jennifer C. Yee
J.C.Y. and I.-G.S. acknowledge support from N.S.F Grant No. AST--2108414.
% Han, Cheongho
Work by C.H. was supported by the grants of National Research Foundation of Korea (2019R1A2C2085965 and 2020R1A4A2002885).
% Yossi Shvartzvald
Y.S. acknowledges support from BSF Grant No. 2020740.
% Weicheng & Hongjing
W.Z. and H.Y. acknowledge support by the National Natural Science Foundation of China (Grant No. 12133005).
W.Z. acknowledges the support from the Harvard-Smithsonian Center for Astrophysics through the CfA Fellowship.
% Hydra
The computations in this paper were conducted on the Smithsonian High Performance Cluster (SI/HPC), Smithsonian Institution (\url{https://doi.org/10.25572/SIHPC}).
% Exoplanet Archive
This research has made use of the NASA Exoplanet Archive, which is operated by the California Institute of Technology, under contract with the National Aeronautics and Space Administration under the Exoplanet Exploration Program.
%\end{acknowledgments}
% =================================

\appendix 

\section{Non-planetary events} \label{sec:appendix_binaries}
From the preliminary analysis using the pipeline data sets, we find that some events in the $2016$ sub-prime fields have the potential to be caused by planetary lens systems (i.e., $q < 0.06$). However, based on the detailed analysis using the TLC data sets, we reveal that these events were caused by binary lens systems. The events cannot satisfy our criteria (i.e., no competing planetary solutions having $q < 0.03$ and $\Delta\chi^{2} < 10.0$). Although the scientific importance is low for these events, we briefly document these binary-lens events for the record. This documentation will be helpful to avoid redundant efforts for planet searches using the KMTNet data archive. In Table \ref{table:obs_planet}, we list these non-planetary events with their observational information.

\subsection{\sixtwenty} % Binary event : KMT-2016-BLG-0255 | OGLE-2016-BLG-0620(*) | MOA-2016-BLG-183
The overall shape of the light curve of \sixtwenty\ is a 1L1S--like feature. However, the 1L1S model exhibits residuals at the rising part of the left wing and around the peak. We find that these systematic residuals can be explained by the 2L1S interpretation, which gives $\Delta\chi^{2} = 613.2$ between 1L1S and 2L1S models. The best--fit 2L1S model has $(s,q) = (2.449 \pm 0.050, 0.208 \pm 0.013)$, which indicates that the lens is a binary system. We also find a planet-like model (i.e., $q = 0.027 \pm 0.004$). However, this case is worse than the best--fit by $\Delta\chi^{2} =30.7$, which does not satisfy our criterion (i.e., $\Delta\chi^{2} < 10.0$). Thus, we conclude that \sixtwenty\ was caused by the binary rather than a planetary system.

\subsection{\ninethirteen} % Binary event : KMT-2016-BLG-0913(*)
The light curve of \ninethirteen\ shows an apparent anomaly at the peak. We find the best--fit 2L1S model has $(s,q) = (2.492 \pm 0.067, 0.823 \pm 0.082)$. The best--fit model indicates that the lens system consists of binary stars. We check possible planetary models. We find two possible cases with $(s,q) = (1.328 \pm 0.032, 0.044 \pm 0.008)$ and $(s,q) = (0.700 \pm 0.009, 0.032 \pm 0.004)$. However, both cases are disfavored by $\Delta\chi^{2} = 104.5$ and $116.7$. Furthermore, the mass ratios of both cases do not satisfy our criterion. Thus, we conclude that \ninethirteen\ was caused by a binary lens system.

\subsection{\fourteenthirtytwo} % Binary event : KMT-2016-BLG-1004 | OGLE-2016-BLG-1432(*)
The light curve of \fourteenthirtytwo\ shows an asymmetric feature. This anomaly can be explained by a 2L1S model with $(s, q) = (1.423 \pm 0.074, 0.257 \pm 0.063)$. The best--fit model indicates the lens is a binary lens system. We also find a planetary model having $(s,q) = (1.603 \pm 0.056, (69.475 \pm 29.048)\times 10^{-4})$. However, the planetary case is disfavored by $\Delta\chi^{2} = 78.1$. Thus, we conclude that \fourteenthirtytwo\ was caused by a binary.

\subsection{\twelvetwentytwo} % Binary event : KMT-2016-BLG-1222(*)
The light curve of \twelvetwentytwo\ shows a clear anomaly at the peak. We find that the 2L1S interpretation can explain the anomaly. The best--fit solution with $(s,q) = (1.926 \pm 0.117, 0.113 \pm 0.024)$ indicates the lens is a binary. We also find a degenerate solution caused by the well-known close/wide degeneracy. This close solution with $(s,q) = (0.584 \pm 0.027, 0.088 \pm 0.017)$ has only $\Delta\chi^{2} = 1.28$, so the degeneracy cannot be resolved. But, the close solution also indicates the lens is a binary. In addition, we find that there is no possible model having $q < 0.03$ based on the detailed analysis using the TLC data sets. Thus, we conclude that \twelvetwentytwo\ was caused by the binary.

\subsection{\eighteenfourtyfour} % Binary event : KMT-2016-BLG-1326 | OGLE-2016-BLG-1844(*)
The light curve of \eighteenfourtyfour\ exhibits two bump-shaped anomalies at the peak. The anomaly can be described by 2L1S models with $(s,q) = (4.062 \pm 0.413, 0.313 \pm 0.088)$ or $(s,q) = (0.337 \pm 0.028, 0.171 \pm 0.034)$ corresponding to the $s_{+}$ or $s_{-}$ cases, respectively. The degeneracy between the $s_{\pm}$ solutions cannot be resolved (i.e., $\Delta\chi^{2} = 0.7$). Both $s_{\pm}$ cases indicate the lens is a binary system. We find possible planetary cases (i.e., $q \sim (38 \pm 5)\times 10^{-4}$) with $s_{+}$ ($s = 1.289 \pm 0.027$) and $s_{-}$ ($s = 0.761 \pm 0.015$), but they are disfavored by $\Delta\chi^{2} = 61.7$ and $62.1$, respectively. Thus, we conclude that \eighteenfourtyfour\ was caused by a binary lens system.

\subsection{\fourteentwentyfive} % Binary event : KMT-2016-BLG-1425
The light curve of \fourteentwentyfive\ shows an apparent bump near the peak. The anomaly can be described by a 2L1S model with $(s, q) = (1.169 \pm 0.057, 0.314 \pm 0.056)$ (the best-fit model). We find an alternative planetary model having $(s,q) = (1.279 \pm 0.062, 0.017 \pm 0.012)$. However, this case is disfavored by $\Delta\chi^{2} = 17.7$, which does not satisfy our criterion (i.e., $\Delta\chi^{2} < 10.0$). Thus, we conclude that \eighteightytwo\ was caused by a binary lens system.

\subsection{\nineeightytwo} % Binary event : KMT-2016-BLG-1433 | OGLE-2016-BLG-0982(*)
The light curve of \nineeightytwo\ shows a clear bump-shaped anomaly near the peak. We find the best-fit solution has $(s,q) = (1.989 \pm 0.086, 0.660 \pm 0.090)$, which indicates the lens is a binary system. There are no possible planetary cases (i.e., $q < 0.03$). Among the competing solutions, the lowest $q$ value is $0.083 \pm 0.011$, which is disfavored by $\Delta\chi^{2} = 16.7$. Thus, we conclude that \nineeightytwo\ was caused by a binary system.

\subsection{\fifteenseventeen} % KMT-2016-BLG-1461
The light curve of \fifteenseventeen\ shows asymmetric deviations from the 1L1S fitting. The anomaly can be explained by the 2L1S models with $(s,q) = (3.259 \pm 0.103, 0.460 \pm 0.083)$. The best-fit solution indicates the lens is a binary. We have checked for possible planetary cases. However, we find that the model having the lowest $q$ value (i.e., $q = 0.014 \pm 0.002$) is disfavored by $\Delta\chi^{2} = 60.0$. Thus, we conclude that \fifteenseventeen\ was caused by a binary lens system.

\subsection{\twelvefiftyeight} % KMT-2016-BLG-2067
\twelvefiftyeight\ is a low-magnification event with a bump at the peak. The anomaly can be described by 2L1S models with $(s,q) = (0.573 \pm 0.012, 0.192 \pm0.031)$ and $(s,q) = (2.002\pm 0.085, 0.161 \pm 0.062)$ corresponding to the $s_{-}$ and $s_{+}$ cases, respectively. The $s_{-}$ case shows better fits than the $s_{+}$ case by $\Delta\chi^{2} = 7.2$. Both solutions imply that the lens is a binary system. We find a possible planetary model ($q = 0.011 \pm 0.003$), which is disfavored by $\Delta\chi^{2} = 27.4$. This planetary case is rejected based on our criterion. Thus, we conclude that \twelvefiftyeight\ was caused by a binary lens system.

\subsection{\twentytwofiftysix} % KMT-2016-BLG-2256
The light curve of \twentytwofiftysix\ exhibits a bump-like anomaly, which is sparsely covered by KMTC only. The anomaly can be explained by both $s_{\pm}$ models with  $(s,q) = (0.580 \pm 0.043, 0.629 \pm 0.197)$ and $(s,q) = (3.937 \pm 0.292, 0.360 \pm 0.410)$ corresponding to the $s_{-}$ and $s_{+}$ cases, respectively. Although the $s_{\pm}$ cases cannot be resolved (i.e., $\Delta\chi^{2} = 1.5$), both cases indicate that the lens is a binary system. We find that there is no competing planetary solution. The lowest $q$ model (i.e., $q = 0.029 \pm 0.005$) is disfavored by $\Delta\chi^{2} = 47.9$, which is clearly rejected based on our criterion. Thus, we conclude that \twentytwofiftysix\ was caused by a binary lens system.

\subsection{\twentythreethirtyone} % KMT-2016-BLG-2331
The light curve of \twentythreethirtyone\ shows a bump at the peak, which is sparsely covered. The anomaly can be described by both 2L1S models with $(s,q) = (0.317 \pm 0.025, 0.280 \pm 0.081)$ and $(s,q) = (5.014 \pm 0.646, 0.478 \pm 0.289)$ corresponding to the $s_{-}$ and $s_{+}$ cases, respectively ($\Delta\chi^{2} =2.7$). Both $s_{\pm}$ cases indicate the lens is a binary. We find a possible planetary model having $q = (92.366 \pm 14.168)\times10^{-4}$. However, this planet case is disfavored by $\Delta\chi^{2} = 57.5$, which is clearly rejected based on our criterion. Thus, we conclude that \twentythreethirtyone\ was caused by a binary lens system.

% References ----------------------------------------------------------------   

\newpage
% =====================================================================
% Tables 
% =====================================================================

% Table X (Observation info: planetary events) ------------------------------
\begin{deluxetable}{ccc|rrr|rc}
\tablecaption{Observations of $2016$ planets and planet candidates \label{table:obs_planet}}
\tablewidth{0pt}
\tablehead{
% ---------------------------------------------------------------------------
\multicolumn{3}{c|}{Event} &
\multicolumn{3}{c|}{Location} &
\multicolumn{2}{c}{obs. info.} \\
% --------------------------------------------
\multicolumn{1}{c}{KMTNet} &
\multicolumn{1}{c}{OGLE} &
\multicolumn{1}{c|}{MOA} &
% --------------------------------------------
\multicolumn{1}{c}{R.A. (J2000)} &
\multicolumn{1}{c}{Dec (J2000)} &
\multicolumn{1}{c|}{$(\ell, b)$} &
% --------------------------------------------
\multicolumn{1}{c}{$A_{I}$} &
\multicolumn{1}{c}{$\Gamma$ (${\rm hr}^{-1}$)} 
% ------------------------------------
}
\startdata
% ---------------------------------------------------------------------------------------------------------------------------------------------
     0696 & \bf{1598} &     521  & $18^{h} 00^{m} 45^{s}.78$ & $-29^{\circ} 10{'} 31{''}.58$ & $(+1^{\circ}.47, -2^{\circ}.97)$ & 1.16 & 1.0 \\
     0781 & \bf{1800} &     581  & $17^{h} 59^{m} 36^{s}.77$ & $-30^{\circ} 51{'} 55{''}.30$ & $(-0^{\circ}.12, -3^{\circ}.59)$ & 1.72 & 1.0 \\
     1611 &     1705  & \bf{526} & $17^{h} 46^{m} 40^{s}.39$ & $-34^{\circ} 23{'} 09{''}.89$ & $(-4^{\circ}.54, -3^{\circ}.02)$ & 1.50 & 0.4 \\
\bf{2321} & \nodata   & \nodata  & $17^{h} 36^{m} 13^{s}.90$ & $-25^{\circ} 31{'} 29{''}.57$ & $(+1^{\circ}.79, +3^{\circ}.58)$ & 3.88 & 0.4 \\
 % ---------------------------------------------------------------------------------------------------------------------------------------------
\hline
 % ---------------------------------------------------------------------------------------------------------------------------------------------
\bf{1243} & \nodata   & \nodata & $17^{h} 49^{m} 55^{s}.56$ & $-22^{\circ} 11{'} 57{''}.91$ & $(+6^{\circ}.27, +2^{\circ}.65)$ & 1.98 & 0.4  \\
    1406  & \bf{0336} & 092     & $18^{h} 16^{m} 09^{s}.86$ & $-25^{\circ} 08{'} 02{''}.29$ & $(+6^{\circ}.68, -4^{\circ}.02)$ & 1.18 & 0.4  \\
    1449  & \bf{0882} & \nodata & $18^{h} 14^{m} 50^{s}.42$ & $-27^{\circ} 40{'} 06{''}.71$ & $(+4^{\circ}.29, -4^{\circ}.96)$ & 0.52 & 0.4  \\
    1609  & \bf{1704} & \nodata & $17^{h} 47^{m} 32^{s}.79$ & $-34^{\circ} 42{'} 40{''}.10$ & $(-4^{\circ}.73, -3^{\circ}.34)$ & 1.04 & 0.4  \\
\enddata
\tablecomments{
The boldface indicates the ``discovery" name of each event. The horizontal lines separate planetary events, planet candidates, and non-planetary events (see Appendix \ref{sec:appendix_binaries}).
}
\end{deluxetable}
% ---------------------------------------------------------------------------

% Table X (Definitions of acronyms and parameters) ------------------------------
\begin{deluxetable}{ll}
\tablecaption{Definition of acronyms and model parameters \label{table:definitions}}
\tablewidth{0pt}
\tablehead{
% ---------------------------------------------------------------------------
\multicolumn{1}{l}{Acronym} & 
\multicolumn{1}{l}{Definition} 
% ------------------------------------
}
\startdata
% ---------------------------------------------------------------------------------------------------------------------------------------------
$n$L$m$S           & Number of lenses ($n$) and sources ($m$), which are included for models                          \\
STD$^{1}$          & Static model without any consideration of acceleration for the lens, source, and observer        \\
APRX               & Model considering the annual microlens parallax (APRX) effect \citep{gould92}                    \\
OBT                & Model considering the effect of the orbital motion (OBT) of lens system                          \\
% ---------------------------------------------------------------------------------------------------------------------------------------------
\hline
\hline
Model parameter    & Definition                                                                                        \\
% ---------------------------------------------------------------------------------------------------------------------------------------------
\hline
$t_0$                      & Time at the peak of the light curve                                                              \\
$u_0$                      & Impact parameter in units of $\theta_{\rm E}$                                                    \\ 
$t_{\rm E}$                & Time during which the source travels the angular Einstein ring radius ($\theta_{\rm E}$)         \\
\enddata
\tablecomments{$^{1}$We conduct modeling using the static case as the standard (STD) model.
}
\end{deluxetable}
% ---------------------------------------------------------------------------

% Table X (KB-16-0696: 2L1S model parameters) --------------------------------
\begin{deluxetable}{lrr}
\tablecaption{The parameters of degenerate 2L1S models for \fifteenninetyeight \label{table:model_0696}}
\tablewidth{0pt}
\tablehead{
% ---------------------------------------------------------------------------
\multicolumn{1}{c}{Parameter} &
\multicolumn{1}{c}{outer} & 
\multicolumn{1}{c}{inner}
% ---------------------------------------------------------------------------
}
\startdata
% -----------------------------------------------------------------------------------
$\chi^{2} / {\rm N}_{\rm data}$  & $ 4393.288 / 4395    $ & $ 4401.751 / 4395    $ \\  
$\Delta\chi^{2}$                 & \nodata (best-fit)     & $ 8.463 $              \\ 
$t_0$ [HJD$^{\prime}$]           & $ 7641.470 \pm 0.037 $ & $ 7641.471 \pm 0.035 $ \\  
$u_0$                            & $    0.225 \pm 0.007 $ & $    0.219 \pm 0.007 $ \\  
$t_{\rm E}$ [days]               & $   37.513 \pm 0.810 $ & $   37.833 \pm 0.814 $ \\  
$s$                              & $    0.955 \pm 0.057 $ & $    0.727 \pm 0.030 $ \\  
$q$ ($\times 10^{-4}$)           & $    6.437 \pm 3.738 $ & $   12.211 \pm 3.543 $ \\  
$\langle\log_{10} q\rangle$      & $   -3.003 \pm 0.152 $ & $   -2.943 \pm 0.131 $ \\  
$\alpha$ [rad]                   & $    5.137 \pm 0.019 $ & $    5.145 \pm 0.016 $ \\  
$\rho_{\ast,{\rm limit}}$        & $  < 0.083           $ & $  < 0.059           $ \\  
% -----------------------------------------------------------------------------------
\enddata
\tablecomments{
HJD$^{\prime} = $HJD$ - 2450000.0$. We note that the $\rho_{\ast}$ is not measured. 
Thus, we present $3\sigma\,$ upper limits on the $\rho_{\ast}$ values 
(i.e., $\rho_{\ast,{\rm limit}}$).
}
\end{deluxetable}
% -----------------------------------------------------------------------------------

% Table X (KB-16-0781: 2L1S model parameters) --------------------------------
\begin{deluxetable}{lrr}
\tablecaption{The parameters of 2L1S solutions for \eighteendoubleO \label{table:model_0781}}
\tablewidth{0pt}
\tablehead{
% ---------------------------------------------------------------------------
\multicolumn{1}{c}{Parameter} &
\multicolumn{1}{c}{$s_{-}$} & 
\multicolumn{1}{c}{$s_{+}$}
% ------------------------------------
}
\startdata
% ------------------------------------------------------------------------------------------------------
$\chi^{2} / {\rm N}_{\rm data}$  & $ 6117.000 / 6145 $     & $ 6117.919 / 6145     $ \\  
$\Delta\chi^{2}$                 & \nodata (best-fit)      & $ 0.919               $ \\ 
$t_0$ [HJD$^{\prime}$]           & $ 7657.025 \pm  0.026 $ & $ 7656.985 \pm  0.030 $ \\  
$u_0$                            & $    0.128 \pm  0.007 $ & $    0.139 \pm  0.009 $ \\  
$t_{\rm E}$ [days]               & $   20.415 \pm  0.847 $ & $   19.844 \pm  0.878 $ \\  
$s$                              & $    0.686 \pm  0.030 $ & $    1.211 \pm  0.066 $ \\  
$q$ ($\times 10^{-4}$)           & $   57.528 \pm 14.341 $ & $   65.310 \pm 23.391 $ \\  
$\langle\log_{10} q\rangle$      & $   -2.229 \pm  0.104 $ & $   -2.133 \pm  0.134 $ \\  
$\alpha$ [rad]                   & $    5.403 \pm  0.016 $ & $    5.392 \pm  0.016 $ \\  
$\rho_{\ast,{\rm limit}}$        & $  < 0.057            $ & $  < 0.071            $ \\  
% ------------------------------------------------------------------------------------------------------
\enddata
\tablecomments{
HJD$^{\prime} = $HJD$- 2450000.0$. We note that the $\rho_{\ast}$ is not measured for both 2L1S cases. 
We present $3\sigma\,$ upper limits on the $\rho_{\ast}$ values (i.e., $\rho_{\ast,{\rm limit}}$).
}
\end{deluxetable}
% ---------------------------------------------------------------------------

% Table X (KB-16-1611: 2L1S model parameters) --------------------------------
\begin{deluxetable}{lrrrrr}
\tablecaption{The parameters of degenerate 2L1S solutions for \fivetwentysix \label{table:model_1611}}
\tablewidth{0pt}
\tablehead{
% ---------------------------------------------------------------------------
\multicolumn{1}{c}{Parameter} &
\multicolumn{1}{c}{$s_{-}$ (A)} & 
\multicolumn{1}{c}{$s_{-}$ (B)} & 
\multicolumn{1}{c}{$s_{-}$ (C)} &
\multicolumn{1}{c}{$s_{+}$ (A)} & 
\multicolumn{1}{c}{$s_{+}$ (B)}
% ------------------------------------
}
\startdata
% --------------------------------------------------------------------------------------------------------------------------------------------------------------
$\chi^{2} / {\rm N}_{\rm data}$  & $ 2218.429 / 2220    $ & $ 2218.431 / 2220    $ & $ 2218.648 / 2220    $ & $ 2219.283 / 2220    $ & $ 2219.463 / 2220    $ \\
$\Delta\chi^{2}$                 & \nodata (best-fit)     & $ 0.002              $ & $ 0.219              $ & $ 0.854              $ & $ 1.034              $ \\
$t_0$ [HJD$^{\prime}$]           & $ 7638.076 \pm 0.021 $ & $ 7638.075 \pm 0.021 $ & $ 7638.078 \pm 0.022 $ & $ 7638.056 \pm 0.019 $ & $ 7638.059 \pm 0.019 $ \\
$u_0$                            & $    0.086 \pm 0.005 $ & $    0.087 \pm 0.005 $ & $    0.090 \pm 0.005 $ & $    0.085 \pm 0.005 $ & $    0.086 \pm 0.005 $ \\
$t_{\rm E}$ [days]               & $   19.483 \pm 0.771 $ & $   19.446 \pm 0.758 $ & $   19.037 \pm 0.774 $ & $   19.558 \pm 0.752 $ & $   19.348 \pm 0.754 $ \\
$s$                              & $    0.944 \pm 0.005 $ & $    0.954 \pm 0.003 $ & $    0.965 \pm 0.006 $ & $    1.062 \pm 0.013 $ & $    1.036 \pm 0.014 $ \\
$q$ ($\times 10^{-4}$)           & $    1.760 \pm 0.467 $ & $    1.184 \pm 0.329 $ & $    1.593 \pm 0.524 $ & $    0.605 \pm 0.620 $ & $    0.475 \pm 0.751 $ \\
$\langle\log_{10} q\rangle$      & $   -3.772 \pm 0.119 $ & $   -3.957 \pm 0.128 $ & $   -3.787 \pm 0.138 $ & $   -4.096 \pm 0.229 $ & $   -4.081 \pm 0.239 $ \\
\enddata
\tablecomments{
HJD$^{\prime} = $ HJD$-2450000.0$. We note that the $\rho_{\ast}$ is not measured for any 2L1S case. 
We present $3\sigma\,$ upper limits on the $\rho_{\ast}$ values (i.e., $\rho_{\ast,{\rm limit}}$).
}
\end{deluxetable}
% ---------------------------------------------------------------------------

% Table X (KB-16-2321: 2L1S model parameters) --------------------------------
\begin{deluxetable}{lrr}
\tablecaption{The parameters of degenerate 2L1S solutions for \twentythreetwentyone \label{table:model_2321}}
\tablewidth{0pt}
\tablehead{
% -----------------------------------------------------------------------------------------------
\multicolumn{1}{c}{Parameter} &
\multicolumn{1}{c}{outer} & 
\multicolumn{1}{c}{inner}
% -----------------------------------------------------------------------------------------------
}
\startdata
% -----------------------------------------------------------------------------------------------
$\chi^{2} / {\rm N}_{\rm data}$              & $ 882.415 / 883      $ & $ 883.079 / 883      $ \\
$\Delta\chi^{2}$                             & \nodata (best-fit)     & $ 0.664              $ \\
$t_0$ [HJD$^{\prime}$]                       & $ 7625.748 \pm 0.210 $ & $ 7625.684 \pm 0.216 $ \\
$u_0$                                        & $    0.166 \pm 0.018 $ & $    0.168 \pm 0.020 $ \\
$t_{\rm E}$ [days]                           & $   56.847 \pm 4.562 $ & $   56.966 \pm 6.387 $ \\
$s$                                          & $    1.039 \pm 0.014 $ & $    1.157 \pm 0.012 $ \\
$q$ ($\times 10^{-4}$)                       & $   12.326 \pm 2.992 $ & $   12.703 \pm 3.200 $ \\
$\langle\log_{10} q\rangle$                  & $   -2.967 \pm 0.115 $ & $   -2.948 \pm 0.118 $ \\
$\alpha$ [rad]                               & $    1.993 \pm 0.020 $ & $    1.981 \pm 0.021 $ \\
$\rho_{\ast,{\rm limit}}$ ($\times 10^{-3}$) & $  < 1.466           $ & $  < 1.268           $ \\
% -----------------------------------------------------------------------------------------------
\enddata
\tablecomments{
HJD$^{\prime} = $ HJD$-2450000.0$. We note that $\rho_{\ast}$ is not measured for any case. 
We present $3\sigma\,$ upper limits on the $\rho_{\ast}$ values (i.e., $\rho_{\ast,{\rm limit}}$).
}
\end{deluxetable}
% -----------------------------------------------------------------------------------------------

% Table X (KB-16-1243: 2L1S model parameters) --------------------------------
\begin{longrotatetable}
\begin{deluxetable}{lrr|rrrr|lr}
\tablecaption{The parameters of degenerate solutions for \twelvefourtythree \label{table:model_1243}}
\tabletypesize{\scriptsize}
\tablewidth{0pt}
\tablehead{
% ---------------------------------------------------------------------------
\multicolumn{1}{c}{Case} &
\multicolumn{2}{c}{2L1S: binary} & 
\multicolumn{4}{|c}{2L1S: planet} & 
\multicolumn{2}{|c}{1L1S} \\
% ------------------------------------
\multicolumn{1}{c}{Parameter} &
\multicolumn{1}{c}{B1} & 
\multicolumn{1}{c}{B2} & 
\multicolumn{1}{|c}{P1} &
\multicolumn{1}{c}{P2} & 
\multicolumn{1}{c}{P3} & 
\multicolumn{1}{c}{P4} & 
\multicolumn{1}{|c}{Parameter} &
\multicolumn{1}{c}{FS}
% ------------------------------------
}
\startdata
% --------------------------------------------------------------------------------------------------------------------------------------------------------------
$\chi^{2} / {\rm N}_{\rm data}$  & $ 676.060 / 677 $      & $ 676.254 / 677      $ & $ 677.904 / 677      $                        & $ 680.261 / 677      $         & $ 680.065 / 677              $ & $ 680.823 / 677              $ & $\chi^{2} / {\rm N}_{\rm data}$ & $ 685.628 / 677      $ \\
$\Delta\chi^{2}$                 & \nodata (best-fit)     & $ 0.194              $ & $ 1.844              $                        & $ 4.201              $         & $ 4.005                      $ & $ 4.763                      $ & $\Delta\chi^{2}$                & $ 9.568              $ \\
$t_0$ [HJD$^{\prime}$]           & $ 7644.621 \pm 0.022 $ & $ 7644.596 \pm 0.020 $ & $ 7644.588 \pm 0.019 $                        & $ 7644.587 \pm 0.019 $         & $ 7644.583 \pm 0.019         $ & $ 7644.598 \pm 0.019         $ & $t_0$ [HJD$^{\prime}$]          & $ 7644.599 \pm 0.018 $ \\
$u_0$                            & $    0.018 \pm 0.001 $ & $    0.015 \pm 0.001 $ & $    0.023 \pm 0.002 $                        & $    0.022 \pm 0.002 $         & $    0.024 \pm 0.002         $ & $    0.024 \pm 0.002         $ & $u_0$                           & $    0.023 \pm 0.002 $ \\
$t_{\rm E}$ [days]               & $   74.153 \pm 4.402 $ & $   85.562 \pm 5.320 $ & $   72.871 \pm 4.421 $                        & $   77.779 \pm 4.802 $         & $   70.841 \pm 4.591         $ & $   71.117 \pm 4.656         $ & $t_{\rm E}$ [days]              & $   74.794 \pm 4.593 $ \\
$s$                              & $    0.162 \pm 0.023 $ & $    8.260 \pm 1.095 $ & $    1.167_{-1.002}^{+0.584}                $ & $    0.407_{-0.248}^{+2.387} $ & $    2.945_{-2.812}^{+2.006} $ & $    0.356_{-0.223}^{+4.462} $ & \nodata                         & \nodata                \\
\enddata
\tabletypesize{\small}
\tablecomments{
HJD$^{\prime} = $ HJD$-2450000.0$. 
For the $\rho_{\ast}$ parameter, the inequality sign indicates the upper limit on $\rho_{\ast}$ (i.e., $3\sigma$), 
because we cannot robustly measure $\rho_{\ast}$ for those cases.
}
%\tabletypesize{\small}
\end{deluxetable}
\end{longrotatetable}
% ---------------------------------------------------------------------------

% Table X (KB-16-1406: 2L1S model parameters) --------------------------------
\begin{deluxetable}{lrrr|lr}
\tablecaption{The parameters of degenerate solutions for \threethirtysix \label{table:model_1406}}
\tablewidth{0pt}
\tablehead{
% ---------------------------------------------------------------------------
\multicolumn{1}{c}{Parameter} &
\multicolumn{1}{c}{A: 2L1S} & 
\multicolumn{1}{c}{B: 2L1S} & 
\multicolumn{1}{c}{C: 2L1S} &
\multicolumn{1}{|c}{Parameter} &
\multicolumn{1}{c}{1L2S}
% ------------------------------------
}
\startdata
% --------------------------------------------------------------------------------------------------------------------------------------------------------------
$\chi^{2} / {\rm N}_{\rm data}$  & $ 7847.419 / 7875    $ & $ 7848.177 / 7875    $ & $ 7848.525 / 7875    $ & $\chi^{2} / {\rm N}_{\rm data}$ & $ 7848.545 / 7875    $ \\
$\Delta\chi^{2}$                 & \nodata (best-fit)     & $ 0.758              $ & $ 1.106              $ & $\Delta\chi^{2}$                & $ 1.126              $ \\
$t_0$ [HJD$^{\prime}$]           & $ 7480.996 \pm 0.009 $ & $ 7481.006 \pm 0.009 $ & $ 7481.009 \pm 0.009 $ & $t_{0,{\rm S1}}$ [HJD$^{\prime}$] & $ 7480.995 \pm 0.010 $ \\
$u_0$                            & $    0.166 \pm 0.002 $ & $    0.164 \pm 0.002 $ & $    0.164 \pm 0.002 $ & $u_{0,{\rm S1}}$                & $    0.168 \pm 0.006 $ \\
$t_{\rm E}$ [days]               & $   24.513 \pm 0.227 $ & $   24.621 \pm 0.230 $ & $   24.623 \pm 0.230 $ & $t_{\rm E}$ [days]              & $   24.495 \pm 0.246 $ \\
$s$                              & $    1.089 \pm 0.003 $ & $    1.007 \pm 0.016 $ & $    1.174 \pm 0.019 $ & $t_{0,{\rm S2}}$ [HJD$^{\prime}$] & $ 7481.700 \pm 0.012 $ \\
$q$ $(\times10^{-4})$            & $    0.229 \pm 0.032 $ & $    1.462 \pm 0.319 $ & $    1.577 \pm 0.340 $ & $u_{0,{\rm S2}}$                & $   -0.004 \pm 0.003 $ \\
$\langle\log_{10} q\rangle$      & $   -4.619 \pm 0.059 $ & $   -3.835 \pm 0.110 $ & $   -3.802 \pm 0.115 $ & $q_{\rm flux}$                  & $    0.002 \pm 0.001 $ \\
\enddata
\tablecomments{
HJD$^{\prime} = $ HJD$ - 2450000.0$. 
For the $\rho_{\ast}$ parameter, the inequality sign indicates the upper limit on $\rho_{\ast}$ (i.e., $3\sigma$), 
because we cannot robustly measure $\rho_{\ast}$ for those cases.
}
\end{deluxetable}
% ---------------------------------------------------------------------------

% Table X (KB-16-1449: 2L2S & 3L1S model parameters) --------------------------------
\begin{deluxetable}{lrrr|lr}
\tablecaption{The parameters of degenerate solutions for \eighteightytwo \label{table:model_1449}}
\tablewidth{0pt}
\tablehead{
% ---------------------------------------------------------------------------
\multicolumn{1}{c}{Parameter} &
\multicolumn{1}{c}{A: 2L2S} & 
\multicolumn{1}{c}{B: 2L2S} & 
\multicolumn{1}{c}{C: 2L2S} &
\multicolumn{1}{|c}{Parameter} &
\multicolumn{1}{c}{3L1S}
% ------------------------------------
}
\startdata
% --------------------------------------------------------------------------------------------------------------------------------------------------------------
$\chi^{2} / {\rm N}_{\rm data}$ & $ 1431.945 / 1436    $ & $ 1439.626 / 1436    $ & $ 1437.736 / 1436    $ & $\chi^{2} / {\rm N}_{\rm data}$ & $ 1471.948 / 1436    $ \\   
$\Delta\chi^{2}$                & \nodata (best-fit)     & $ 7.681              $ & $ 5.791              $ & $\Delta\chi^{2}$                & $ 40.003             $ \\
$t_{0,{\rm S1}}$ [HJD$^{\prime}$] & $ 7520.424 \pm 0.152 $ & $ 7521.433 \pm 0.284 $ & $ 7521.529 \pm 0.035 $ & $t_{0}$                         & $ 7524.747 \pm 0.133 $ \\ 
$u_{0,{\rm S1}}$                & $    0.086 \pm 0.007 $ & $    0.064 \pm 0.006 $ & $    0.350 \pm 0.021 $ & $u_{0}$                         & $    0.095 \pm 0.007 $ \\ 
$t_{\rm E}$ [days]              & $   38.718 \pm 2.156 $ & $   55.660 \pm 4.965 $ & $   32.083 \pm 0.923 $ & $t_{\rm E}$                     & $   54.345 \pm 2.877 $ \\ 
$s$                             & $    0.543 \pm 0.012 $ & $    2.651 \pm 0.269 $ & $    1.186 \pm 0.012 $ & $s_{1}$                         & $    1.523 \pm 0.033 $ \\ 
$q$                             & $    0.286 \pm 0.017 $ & $    0.568 \pm 0.148 $ & $    0.010 \pm 0.001 $ & $q_{1}$                         & $    0.074 \pm 0.005 $ \\ 
$<\log_{10}(q)>$                & $   -0.530 \pm 0.024 $ & $   -0.274 \pm 0.111 $ & $   -1.976 \pm 0.042 $ & $<\log_{10}(q_{1})>$            & $   -1.156 \pm 0.029 $ \\
\enddata
\tablecomments{
HJD$^{\prime} = $ HJD$ - 2450000.0$. 
For the $\rho_{\ast,{\rm S1}}$ and $\rho_{\ast,{\rm S2}}$ parameter, the cases with inequality signs are upper limits (i.e., $3\sigma$), because we cannot robustly measure the source sizes.}
\end{deluxetable}
% ---------------------------------------------------------------------------

% Table X (KB-16-1609: 2L1S model parameters) --------------------------------
%\begin{longrotatetable}
\begin{deluxetable}{lrr|rr|lr}
\tablecaption{The parameters of degenerate solutions for \seventeenfour \label{table:model_1609}}
%\tabletypesize{\scriptsize}
\tablewidth{0pt}
\tablehead{
% ---------------------------------------------------------------------------
\multicolumn{1}{c}{Case} &
\multicolumn{2}{c}{2L1S: binary} & 
\multicolumn{2}{|c}{2L1S: planet-like} & 
\multicolumn{2}{|c}{1L2S} \\
% ------------------------------------
\multicolumn{1}{c}{Parameter} &
\multicolumn{1}{c}{(A): $s_{-}$} & 
\multicolumn{1}{c}{(B): $s_{+}$} & 
\multicolumn{1}{|c}{(C): $s_{-}$} &
\multicolumn{1}{c}{(D): $s_{+}$} & 
\multicolumn{1}{|c}{Parameter} &
\multicolumn{1}{c}{(E)}
% ------------------------------------
}
\startdata
% --------------------------------------------------------------------------------------------------------------------------------------------------------------
$\chi^{2} / {\rm N}_{\rm data}$  & $ 1791.049 / 1796    $ & $ 1791.171 / 1796    $ & $ 1793.005 / 1796    $ & $ 1804.127 / 1796    $ & $\chi^{2} / {\rm N}_{\rm data}$ & $ 1794.442 / 1796    $ \\
$\Delta\chi^{2}$                 & \nodata (best-fit)     & $ 0.122              $ & $ 1.956              $ & $ 13.078             $ & $\Delta\chi^{2}$                & $ 3.393              $ \\
$t_0$ [HJD$^{\prime}$]             & $ 7638.124 \pm 0.198 $ & $ 7638.611 \pm 0.086 $ & $ 7638.207 \pm 0.107 $ & $ 7638.209 \pm 0.090 $ & $t_{0,{\rm S1}}$ [HJD$^{\prime}$] & $ 7637.907 \pm 0.193 $ \\
$u_0$                            & $    0.125 \pm 0.023 $ & $    0.028 \pm 0.010 $ & $    0.155 \pm 0.023 $ & $    0.072 \pm 0.011 $ & $u_{0,{\rm S1}}$                & $    0.238 \pm 0.069 $ \\
$t_{\rm E}$ [days]               & $   39.040 \pm 4.575 $ & $   40.442 \pm 6.261 $ & $   32.127 \pm 3.267 $ & $   50.567 \pm 7.298 $ & $t_{\rm E}$ [days]              & $   33.855 \pm 4.459 $ \\
$s$                              & $    0.363 \pm 0.054 $ & $    5.792 \pm 0.443 $ & $    0.723 \pm 0.066 $ & $    1.481 \pm 0.051 $ & $t_{0,{\rm S2}}$ [HJD$^{\prime}$] & $ 7639.517 \pm 0.061 $ \\
$q$                              & $    0.530 \pm 0.240 $ & $    0.076 \pm 0.067 $ & $    0.025 \pm 0.016 $ & $    0.020 \pm 0.005 $ & $u_{0,{\rm S2}}$                & $   -0.006 \pm 0.025 $ \\
$\langle\log_{10} q\rangle$      & $   -0.352 \pm 0.202 $ & $   -1.121 \pm 0.283 $ & $   -1.474 \pm 0.177 $ & $   -1.702 \pm 0.112 $ & $q_{\rm flux}$                  & $    0.059 \pm 0.033 $ \\
\enddata
%\tabletypesize{\small}
\tablecomments{
HJD$^{\prime} = $ HJD$ - 2450000.0$. 
We note that $\rho_{\ast}$ is not measured for all cases. 
Thus, we present $3\sigma\,$ upper limits on the $\rho_{\ast}$ values (i.e., $\rho_{\ast,{\rm limit}}$).
}
%\tabletypesize{\small}
\end{deluxetable}
%\end{longrotatetable}
% ---------------------------------------------------------------------------

% Table X (KB-16-1630: 2L1S APRX model parameters) --------------------------------
%\begin{longrotatetable}
\begin{deluxetable}{lrrr|rr}
\tablecaption{The parameters of 2L1S APRX models for \fourteeneight \label{table:model_1630}}
%\tabletypesize{\scriptsize}
\tablewidth{0pt}
\tablehead{
% ---------------------------------------------------------------------------
\multicolumn{1}{c}{Case} &
\multicolumn{3}{c}{Planet} &
\multicolumn{2}{|c}{Binary} \\
% ---------------------------------------------------------------------------
\multicolumn{1}{c}{Parameter} &
\multicolumn{1}{r}{Resonant ($u_{0}-$)} & 
\multicolumn{1}{r}{Close ($u_{0}-$)} & 
\multicolumn{1}{r}{Wide ($u_{0}-$)} &
\multicolumn{1}{|r}{Close ($u_{0}-$)} &
\multicolumn{1}{r}{Wide ($u_{0}-$)}
% ------------------------------------
}
\startdata
% --------------------------------------------------------------------------------------------------------------------------------------------------------------
$\chi^{2} / {\rm N}_{\rm data}$  & $ 2109.666 / 2124    $ & $ 2119.292 / 2124     $ & $ 2121.676 / 2124     $ & $ 2110.571 / 2124    $ & $ 2111.403 / 2124    $ \\
$\Delta\chi^{2}$                 & \nodata (best-fit)     & $ 9.626               $ & $ 12.010              $ & $ 0.905              $ & $ 1.737              $ \\
$t_0$ [HJD$^{\prime}$]             & $ 7653.474 \pm 0.048 $ & $ 7653.458 \pm  0.051 $ & $ 7653.432 \pm  0.055 $ & $ 7653.763 \pm 0.103 $ & $ 7653.127 \pm 0.070 $ \\
$u_0$                            & $   -0.089 \pm 0.002 $ & $   -0.089 \pm  0.002 $ & $   -0.090 \pm  0.003 $ & $   -0.090 \pm 0.003 $ & $   -0.073 \pm 0.005 $ \\
$t_{\rm E}$ [days]               & $   96.686 \pm 2.217 $ & $   96.302 \pm  2.287 $ & $   95.185 \pm  2.331 $ & $   95.874 \pm 2.872 $ & $  117.147 \pm 8.281 $ \\
$s$                              & $    1.083 \pm 0.005 $ & $    0.701 \pm  0.094 $ & $    1.742 \pm  0.235 $ & $    0.263 \pm 0.018 $ & $    5.400 \pm 0.586 $ \\
$q$ ($\times 10^{-4}$)           & $   18.382 \pm 2.461 $ & $   47.894 \pm 40.051 $ & $   65.010 \pm 37.037 $ & \nodata                & \nodata                \\
$q$                              & \nodata                & \nodata                 & \nodata                 & $    0.275 \pm 0.071 $ & $    0.571 \pm 0.199 $ \\
\enddata
%\tabletypesize{\small}
\tablecomments{
${\rm HJD^{\prime} = HJD - 2450000.0}$. 
For the $\rho_{\ast}$ parameter, the cases with the inequality signs are upper limits on the $\rho_{\ast}$ (i.e., $3\sigma$) 
because we cannot robustly measure $\rho_{\ast}$.
}
\end{deluxetable}
%\end{longrotatetable}
% ---------------------------------------------------------------------------

% Table X (KB-16-2399: 2L1S model parameters) --------------------------------
\begin{deluxetable}{lrrr|lr}
\tablecaption{The parameters of 2L1S and 1L2S models for \twentythreeninetynine \label{table:model_2399}}
\tablewidth{0pt}
\tablehead{
% ---------------------------------------------------------------------------
\multicolumn{1}{c}{Parameter} &
\multicolumn{1}{c}{A: 2L1S} & 
\multicolumn{1}{c}{B: 2L1S} & 
\multicolumn{1}{c}{C: 2L1S} &
\multicolumn{1}{|c}{Parameter} &
\multicolumn{1}{c}{D: 1L2S}
% ------------------------------------
}
\startdata
% --------------------------------------------------------------------------------------------------------------------------------------------------------------
$\chi^{2} / {\rm N}_{\rm data}$  & $ 630.708 / 632      $ & $ 640.907 / 632      $ & $ 645.397 / 632      $ & $\chi^{2} / {\rm N}_{\rm data}$ & $ 644.879 / 632      $ \\
$\Delta\chi^{2}$                 & \nodata (best-fit)     & $ 10.199             $ & $ 14.689             $ & $\Delta\chi^{2}$                & $ 14.171             $ \\
$t_0$ [HJD$^{\prime}$]             & $ 7633.696 \pm 0.088 $ & $ 7633.798 \pm 0.067 $ & $ 7633.450 \pm 0.090 $ & $t_{0,{\rm S1}}$ [HJD$^{\prime}$] & $ 7625.980 \pm 0.097 $ \\
$u_0$                            & $    0.084 \pm 0.007 $ & $    0.173 \pm 0.020 $ & $    0.462 \pm 0.061 $ & $u_{0,{\rm S1}}$                & $    0.002 \pm 0.085 $ \\
$t_{\rm E}$ [days]               & $   19.173 \pm 1.219 $ & $   13.168 \pm 1.265 $ & $    7.970 \pm 0.587 $ & $t_{\rm E}$ [days]              & $   11.171 \pm 1.991 $ \\
$s$                              & $    1.229 \pm 0.016 $ & $    1.646 \pm 0.062 $ & $    0.607 \pm 0.024 $ & $t_{0,{\rm S2}}$ [HJD$^{\prime}$] & $ 7634.150 \pm 0.068 $ \\
$q$                              & $    0.057 \pm 0.009 $ & $    0.030 \pm 0.006 $ & $    0.049 \pm 0.008 $ & $u_{0,{\rm S2}}$                & $    0.221 \pm 0.150 $ \\
$\langle\log_{10} q\rangle$      & $   -1.227 \pm 0.068 $ & $   -1.580 \pm 0.104 $ & $   -1.303 \pm 0.069 $ & $q_{\rm flux}$                  & $   12.943 \pm 2.610 $ \\
\enddata
\tablecomments{
HJD$^{\prime} = $ HJD$ - 2450000.0$. 
For the $\rho_{\ast}$ parameter, the cases with the inequality signs are upper limits on the $\rho_{\ast}$ (i.e., $3\sigma$) 
because we cannot robustly measure $\rho_{\ast}$.
}
\end{deluxetable}
% ---------------------------------------------------------------------------

% Table X (KB-16-2473: 2L1S and 1L2S model parameters) --------------------------------
\begin{deluxetable}{lr|lr}
\tablecaption{The parameters of 2L1S and 1L2S models for \twentyfourseventythree \label{table:model_2473}}
\tablewidth{0pt}
\tablehead{
% ---------------------------------------------------------------------------
\multicolumn{1}{c}{Parameter} &
\multicolumn{1}{c}{2L1S} & 
\multicolumn{1}{|c}{Parameter} &
\multicolumn{1}{c}{1L2S}
% ------------------------------------
}
\startdata
% --------------------------------------------------------------------------------------------------------------------------------------------------------------
$\chi^{2} / {\rm N}_{\rm data}$  & $ 1529.729 / 1535    $ & $\chi^{2} / {\rm N}_{\rm data}$ & $ 1540.057 / 1535     $ \\
$\Delta\chi^{2}$                 & \nodata (best-fit)     & $\Delta\chi^{2}$                & $ 10.328              $ \\
$t_0$ [HJD$^{\prime}$]             & $ 7491.437 \pm 0.206 $ & $t_{0,{\rm S1}}$ [HJD$^{\prime}$] & $ 7515.409 \pm  0.322 $ \\
$u_0$                            & $    0.088 \pm 0.009 $ & $u_{0,{\rm S1}}$                & $    0.033 \pm  0.036 $ \\
$t_{\rm E}$ [days]               & $   47.179 \pm 4.634 $ & $t_{\rm E}$ [days]              & $   51.742 \pm 11.156 $ \\
$s$                              & $    1.217 \pm 0.026 $ & $t_{0,{\rm S2}}$ [HJD$^{\prime}$] & $ 7490.605 \pm  0.160 $ \\
$q$                              & $    0.011 \pm 0.003 $ & $u_{0,{\rm S2}}$                & $    0.103 \pm  0.043 $ \\
$\langle\log_{10} q\rangle$      & $   -1.967 \pm 0.137 $ & $q_{\rm flux}$                  & $    7.663 \pm  1.829 $ \\
\enddata
\tablecomments{
HJD$^{\prime} = $ HJD$ - 2450000.0$. 
For the $\rho_{\ast}$ parameters, the cases with the inequality signs are upper limits on the $\rho_{\ast}$ (i.e., $3\sigma$) 
because we cannot robustly measure $\rho_{\ast}$.
}
\end{deluxetable}
% ---------------------------------------------------------------------------

% Table X (source color) ----------------------------------------------------
%\begin{longrotatetable}
\begin{deluxetable}{crrrrrrrr}
\tablecaption{CMD analyses of Planetary Events \label{table:cmd}}
\tablewidth{0pt}
\tabletypesize{\scriptsize}
\tablehead{
% ---------------------------------------------------------------------------
\multicolumn{1}{c}{Event} & 
\multicolumn{1}{c}{} & 
\multicolumn{1}{c}{} & 
\multicolumn{1}{c}{} & 
\multicolumn{1}{c}{} & 
\multicolumn{1}{c}{} & 
\multicolumn{1}{c}{} &
\multicolumn{1}{c}{} &
\multicolumn{1}{c}{} \\
% ----------------------------------------------
\multicolumn{1}{c}{Case} &
\multicolumn{1}{r}{$(V-I)_{\rm RGC}$} & 
\multicolumn{1}{r}{$(V-I)_{0, \rm RGC}$} &
\multicolumn{1}{r}{$(V-I)_{\rm S}$} &
\multicolumn{1}{r}{$(V-I)_{0, \rm S}$} &
\multicolumn{1}{r}{$(V-I)_{\rm B}$} &
\multicolumn{1}{c}{$\theta_{\ast}$} &
\multicolumn{1}{c}{$\theta_{\rm E}$} &
\multicolumn{1}{c}{$\mu_{\rm rel}$} \\
% ----------------------------------------------
\multicolumn{1}{c}{} &
\multicolumn{1}{r}{$I_{\rm RGC}$} & 
\multicolumn{1}{r}{$I_{0, \rm RGC}$} &
\multicolumn{1}{r}{$I_{\rm S}$} &
\multicolumn{1}{r}{$I_{0, \rm S}$} &
\multicolumn{1}{r}{$I_{\rm B}$} &
\multicolumn{1}{c}{($\mu{\rm as}$)} &
\multicolumn{1}{c}{(mas)} &
\multicolumn{1}{c}{($\rm mas\, yr^{-1}$)}
% ------------------------------------
}
\startdata
% ------------------------------------------------------------------------------------------------------------------------------------------------------------
% KB160696 == OB161598(*)
OB161598    &            &            &                      &                      &                      &                     &             &             \\
outer       & $  2.012 $ & $  1.060 $ & $  1.785 \pm 0.022 $ & $  0.833 \pm 0.055 $ & $  1.161 \pm 0.008 $ & $ 1.080 \pm 0.065 $ & $ > 0.013 $ & $ > 0.127 $ \\
            & $ 15.648 $ & $ 14.385 $ & $ 18.886 \pm 0.006 $ & $ 17.623 \pm 0.006 $ & $ 17.743 \pm 0.004 $ & \nodata             & \nodata     & \nodata     \\
inner       & $  2.012 $ & $  1.060 $ & $  1.784 \pm 0.022 $ & $  0.832 \pm 0.055 $ & $  1.165 \pm 0.008 $ & $ 1.068 \pm 0.065 $ & $ > 0.018 $ & $ > 0.175 $ \\
            & $ 15.648 $ & $ 14.385 $ & $ 18.907 \pm 0.006 $ & $ 17.645 \pm 0.006 $ & $ 17.736 \pm 0.004 $ & \nodata             & \nodata     & \nodata     \\
% ------------------------------------------------------------------------------------------------------------------------------------------------------------
\hline      
% ------------------------------------------------------------------------------------------------------------------------------------------------------------
% KB160781 == OB161800(*) == MB16581
OB161800    &            &            &                      &                      &                      &                     &             &             \\
$s_{-}$     & $  2.489 $ & $  1.060 $ & $  1.991 \pm 0.058 $ & $  0.562 \pm 0.076 $ & \nodata              & $ 0.591 \pm 0.046 $ & $ > 0.010 $ & $ > 0.185 $ \\
            & $ 16.296 $ & $ 14.450 $ & $ 20.139 \pm 0.012 $ & $ 18.294 \pm 0.012 $ & \nodata              & \nodata             & \nodata     & \nodata     \\
\enddata
\tablecomments{We use the abbreviation for event names, e.g., \fifteenninetyeight\ is abbreviated as OB161598.
$^{\ast}$For KB162321, we note that the $V$ and $I$ magnitudes are in units of the instrumental scale of the KMTNet. 
Because there is no available OGLE-III catalog for this event, the magnitude system is not scaled to the OGLE-III. 
}
\tabletypesize{\small}
\end{deluxetable}
%\end{longrotatetable}
% ------------------------------------------------------------------------------------------------------------------------------------------------------------

% Table X (Lens properties) -------------------------------------------------
%\begin{longrotatetable}
\begin{deluxetable}{lccrrrrrcc}
\tablecaption{Lens Properties of Planetary Events \label{table:lens}}
%\tabletypesize{\scriptsize}
\tablewidth{0pt}
\tablehead{
% ---------------------------------------------------------------------------
\multicolumn{1}{c}{Event}                 &
\multicolumn{1}{c}{Constraints}           &
\multicolumn{1}{c}{Case}                  &
\multicolumn{1}{c}{$M_{\rm host}$}        &
\multicolumn{1}{c}{$M_{\rm planet}$}      &
\multicolumn{1}{c}{$D_{\rm L}$}           &
\multicolumn{1}{c}{$a_{\perp}$}           &
\multicolumn{1}{c}{$\mu_{\rm rel}$}       &
\multicolumn{1}{c}{Gal. Mod.}             &
\multicolumn{1}{c}{$\chi^{2}$}            \\
% ---------------------------------------------------------------------------
\multicolumn{1}{c}{}                      &
\multicolumn{1}{c}{}                      &
\multicolumn{1}{c}{}                      &
\multicolumn{1}{c}{($M_{\odot}$)}         &
\multicolumn{1}{c}{($M_{\rm J}$ / $M_{\rm N}$ / $M_{\oplus}$)} &
\multicolumn{1}{c}{(kpc)}                 &
\multicolumn{1}{c}{(au)}                  &
\multicolumn{1}{c}{($\rm mas\, yr^{-1}$)} &
\multicolumn{1}{c}{}                      &
\multicolumn{1}{c}{}                      
% ------------------------------------
}
\startdata
% --------------------------------------------------------------------------------------------------------------------------------------------------------------------------------------------------------------------------------
% KB160696 == OB161598(*)
OB161598      & $t_{\rm E} + \rho_{\ast}$    & outer       & $ 0.55_{-0.32}^{+0.34} $ & $ 0.37_{-0.30}^{+0.31}\, M_{\rm J} $ & $ 5.91_{-2.27}^{+1.28} $ & $ 2.51_{-0.92}^{+0.89} $ & $ 4.65_{-1.75}^{+2.76} $ & 0.714 & 1.000 \\
              &                              & inner       & $ 0.55_{-0.32}^{+0.34} $ & $ 0.70_{-0.46}^{+0.48}\, M_{\rm J} $ & $ 5.91_{-2.27}^{+1.28} $ & $ 1.92_{-0.70}^{+0.67} $ & $ 4.63_{-1.74}^{+2.77} $ & 1.000 & 0.015 \\
{\bf Adopted} &                              &             & $ 0.55 \pm 0.32 $        & $ 0.38 \pm 0.30\, M_{\rm J}$         & $ 5.91 \pm 1.74 $        & $ 2.50 \pm 0.88 $        & $ 4.65 \pm 2.21 $        &       &       \\
% --------------------------------------------------------------------------------------------------------------------------------------------------------------------------------------------------------------------------------
\hline
% --------------------------------------------------------------------------------------------------------------------------------------------------------------------------------------------------------------------------------
% KB160781 == OB161800(*) == MB16581
OB161800      & $t_{\rm E} + \rho_{\ast}$    & $s_{-}$     & $ 0.41_{-0.26}^{+0.33} $ & $ 2.49_{-1.65}^{+2.13}\, M_{\rm J} $ & $ 6.47_{-1.65}^{+1.14} $ & $ 1.49_{-0.56}^{+0.59} $ & $ 6.36_{-2.33}^{+2.90} $ & 1.000 & 1.000 \\
              &                              & $s_{+}$     & $ 0.40_{-0.26}^{+0.33} $ & $ 2.77_{-1.83}^{+2.67}\, M_{\rm J} $ & $ 6.49_{-1.63}^{+1.14} $ & $ 2.60_{-0.97}^{+1.05} $ & $ 6.43_{-2.35}^{+2.93} $ & 0.850 & 0.632 \\
\enddata
\tablecomments{For the planet mass, we present the value in Jupiter ($M_{\rm J}$), Neptune ($M_{\rm N}$), or Earth ($M_{\oplus}$) masses as appropriate.  
}
%\tabletypesize{\small}
\end{deluxetable}
%\end{longrotatetable}
% --------------------------------------------------------------------------- 

% Table X (2016 Planetary events) --------------------------------------------
\begin{deluxetable}{lllccccl}
\tablecaption{Planetary events discovered or recovered by the KMTNet AnomalyFinder in $2016$ \label{table:2016_planets}}
\tabletypesize{\scriptsize}
\tablewidth{0pt}
\tablehead{
% ---------------------------------------------------------------------------
\multicolumn{1}{c}{Event Name}     &
\multicolumn{1}{c}{KMT Name}       &
\multicolumn{1}{c}{KMT Field}      &
\multicolumn{1}{c}{$\log_{10}(q)$} &
\multicolumn{1}{c}{$s$}            &
\multicolumn{1}{c}{Degeneracy}     &
\multicolumn{1}{c}{Method}         &
\multicolumn{1}{c}{Reference}             
% ---------------------------------------------------------------------------
}
\startdata
% -------------------------------------------------------------------------------------------------------------------------------------
KB161105                  &          & sub-prime & -5.19 & 1.14 & i/o, c/w                    & AF     & \citet{zang23}      \\
OB160007                  & KB161991 & prime     & -5.17 & 2.83 &                             & AF     & \citet{zangyy}      \\
OB161195$^{\ast}$         & KB160372 & prime     & -4.34 & 0.99 & c/w, ecliptic               & by-eye & \citet{gould23}     \\
OB161850                  & KB161307 & prime     & -4.00 & 0.80 & i/o, ecliptic               & AF     & \citet{shin23b}     \\
OB161598                  & KB160696 & sub-prime & -3.19 & 0.96 & i/o                         & AF     & This work           \\
KB162321                  &          & sub-prime & -2.91 & 1.04 & i/o                         & AF     & This work           \\
OB161067                  & KB161453 & sub-prime & -2.84 & 0.81 & $s$--degen., ecliptic       & by-eye & \citet{calchi19}    \\
OB161093                  & KB161345 & sub-prime & -2.84 & 1.02 & ecliptic                    & by-eye & \cite{shin22}       \\
\enddata
\tablecomments{
The horizontal line separates planets expected to be part of the final statistical sample and those whose mass ratios are likely too uncertain or too large to be included.
$^{\ast}$For OB161195, the properties of this planetary system was reported by \citet{shvartzvald17} and \citet{bond17}. 
However, we adopt $\log_{10}(q)$ and $s$ values from \citet{gould23}, which re-analyze the event and measure 
a more precise mass ratio. 
$^{\ast\ast}$For OB160613, the event was caused by a lens system consisting of a planet and binary host stars.
$^{\ast\ast\ast}$For KB161855 and KB162142, these are planet candidates.
In the column of ``Degeneracy'', we present the type of degeneracies for the solutions: 
``c/w'', ``i/o'', ``ecliptic'', ``offset'', ``$\alpha$'', and ``1L2S'' indicate
the close/wide degeneracy, inner/outer degeneracy, ecliptic degeneracy of the microlens--parallax effect,
offset--degeneracy, $\alpha$--degeneracy \citep[see][]{shin23a}, and 2L1S/1L2S degeneracy, 
respectively.
Note that ``$s$--degen." indicates small/large $s$ degeneracy \citep[this is different from ``c/w", see][]{calchi19}.
}
\tabletypesize{\small}
\end{deluxetable}
% --------------------------------------------------------------------------- 

\newpage
% =====================================================================
% Figures
% =====================================================================

% Figure X (KB-16-00696: Light curves 2L1S vs. 1L1S) -----------------------------------------------
% --------------------------------------------------------------------------------------------------

% Figure X (KB-16-00696: APRX Test) ----------------------------------------------
% --------------------------------------------------------------------------------------------------

% Figure X (KB-16-0781: Light curves) -----------------------------------------------
% --------------------------------------------------------------------------------------------------

% Figure X (KB-16-1611: Light curves 2L1S vs. 1L1S) -----------------------------------------------
% --------------------------------------------------------------------------------------------------

% Figure X (KB-16-1611: Light curves 2L1S close cases) -----------------------------------------------
% --------------------------------------------------------------------------------------------------

% Figure X (KB-16-1611: Light curves 2L1S wide cases) -----------------------------------------------
% --------------------------------------------------------------------------------------------------

% Figure X (KB-16-2321: Light curves 2L1S vs. 1L1S) -----------------------------------------------
% --------------------------------------------------------------------------------------------------

% Figure X (KB-16-1243: Light curves compared 2L1S: binary, 2L1S: planet, and 1L1S: FS) ------------
% --------------------------------------------------------------------------------------------------

% Figure X (KB-16-1243: Residuals and geometries) --------------------------------------------------
% --------------------------------------------------------------------------------------------------

% Figure X (KB-16-1406: Light curves of degenerate models with xi vs. log(q) space and caustic geometry) ------------
% ---------------------------------------------------------------------------------------------------------------------

% Figure X (KB-16-1449: Light curves of degenerate models with caustic geometry) --------------------------------------
% ---------------------------------------------------------------------------------------------------------------------

% Figure X (KB-16-1609: Light curves of degenerate models with caustic geometry) --------------------------------------
% ---------------------------------------------------------------------------------------------------------------------

% Figure X (KB-16-1630: Light curves of STD and APRX models) ----------------------------------------------------------
% ---------------------------------------------------------------------------------------------------------------------

% Figure X (KB-16-1630: Light curves of degenerate 2L1S APRX models) ----------------------------------------------------------
% ---------------------------------------------------------------------------------------------------------------------

% Figure X (KB-16-1630: Xallarap Test) ----------------------------------------------------------
% ---------------------------------------------------------------------------------------------------------------------

% Figure X (KB-16-2399: Light curves of 2L1S and 1L2S models) ----------------------------------------------------------
% ---------------------------------------------------------------------------------------------------------------------

% Figure X (KB-16-2473: Light curves of 2L1S and 1L2S models) ----------------------------------------------------------
% ---------------------------------------------------------------------------------------------------------------------

% Figure X (CMDs) ---------------------------------------------------------------------------------
% --------------------------------------------------------------------------------------------------

% Figure X (cumulative number of planets) ----------------------------------------------------------
% --------------------------------------------------------------------------------------------------

\end{document}